\RequirePackage{snapshot}
\nonstopmode

\documentclass{jfm}

\usepackage[intlimits]{amsmath}
\RequirePackage{snapshot}
\usepackage{setspace}
\usepackage{graphicx}
\usepackage{pgfplots}
\usepgfplotslibrary{groupplots}
\usepackage{natbib}
\usepackage{overpic}
\usepackage{color}
\usepackage{bibentry}
\usepackage[mathscr]{euscript}
\usepackage{setspace}
\usepackage{arydshln}
\usepackage{tikz}
\usepackage{cases}
\usepackage[overload]{empheq}
\usepackage{cleveref}
\usepackage{url}
\usepackage{wasysym}

\usepackage[utf8]{inputenc}
\usepackage[T1]{fontenc}

\definecolor{BLUE}{rgb}{0.00, 0.00, 0.66}
\definecolor{RED}{rgb}{0.86, 0.00, 0.00}
\definecolor{LRED}{rgb}{0.89, 0.19, 0.19}

\ifCUPmtlplainloaded \else
  \checkfont{eurm10}
  \iffontfound
    \IfFileExists{upmath.sty}
      {\typeout{^^JFound AMS Euler Roman fonts on the system,
                   using the 'upmath' package.^^J}%
       \usepackage{upmath}}
      {\typeout{^^JFound AMS Euler Roman fonts on the system, but you
                   dont seem to have the}%
       \typeout{'upmath' package installed. JFM.cls can take advantage
                 of these fonts,^^Jif you use 'upmath' package.^^J}%
      }
  \else
  \fi
\fi

\makeatletter
\newsavebox{\@brx}
\newcommand{\llangle}[1][]{\savebox{\@brx}{\(\m@th{#1\langle}\)}%
  \mathopen{\copy\@brx\kern-0.5\wd\@brx\usebox{\@brx}}}
\newcommand{\rrangle}[1][]{\savebox{\@brx}{\(\m@th{#1\rangle}\)}%
  \mathclose{\copy\@brx\kern-0.5\wd\@brx\usebox{\@brx}}}
\makeatother

\makeatletter
\newcommand{\specialnumber}[1]{%
  \def\tagform@##1{\maketag@@@{(\ignorespaces##1\unskip\@@italiccorr\emph{#1})}}%
}
\newcommand{\specialeqref}[2]{\begingroup
  \def\tagform@##1{\maketag@@@{(\ignorespaces##1\unskip\@@italiccorr\emph{#2})}}%
  \eqref{#1}\endgroup}
\makeatother

\ifCUPmtlplainloaded \else
  \checkfont{msam10}
  \iffontfound
    \IfFileExists{amssymb.sty}
      {\typeout{^^JFound AMS Symbol fonts on the system, using the
                'amssymb' package.^^J}%
       \usepackage{amssymb}%
         \let\leq=\leqslant
         \let\geq=\geqslant
      }{}
  \fi
\fi

\ifCUPmtlplainloaded \else
  \IfFileExists{amsbsy.sty}
    {\typeout{^^JFound the 'amsbsy' package on the system, using it.^^J}%
     \usepackage{amsbsy}}
    {\providecommand\boldsymbol[1]{\mbox{\boldmath $##1$}}}
\fi

\title[The entrainment and energetics of turbulent plumes in a confined space]{The entrainment and energetics of turbulent plumes in a confined space} \author[John Craske and Megan S. Davies Wykes]%
{John Craske$^1$\thanks{Email address for correspondence:
    john.craske07@imperial.ac.uk} and Megan S. Davies Wykes$^2$}
\affiliation{$^1$Department of Civil and Environmental
  Engineering, Imperial College London,\\[\affilskip] London SW7 2AZ, UK
\\[\affilskip] $^2$ Engineering Department, University
of Cambridge, Trumpington Street,\\[\affilskip]  Cambridge CB2 1PZ, UK}
\pubyear{}
\volume{}
\pagerange{}
\date{?; revised ?; accepted ?. - To be entered by editorial office}

\newcommand{\rd}{\mathrm{d}}
\newcommand{\av}[1]{\overline{#1}}
\newcommand{\hav}[1]{\langle{#1}\rangle}
\newcommand{\vav}[1]{\llangle{#1}\rrangle}

\newcommand{\od}[2]{\dfrac{\mathrm{d} #1}{\mathrm{d} #2}}
\newcommand{\ld}[2]{\dfrac{D #1}{D #2}}
\newcommand{\pd}[2]{\dfrac{\partial #1}{\partial #2}}

\newcommand{\vc}[1]{\boldsymbol{#1}}
\newcommand{\ez}{\boldsymbol{k}}

\begin{document}

\maketitle

\begin{abstract}

We analyse the entrainment and energetics of equal and opposite axisymmetric turbulent air plumes in a vertically confined space at a Rayleigh number of $1.24\times 10^{7}$ using theory and direct numerical simulation. On domains of sufficiently large aspect ratio, the steady-state consists of turbulent plumes penetrating an interface between two layers of approximately uniform buoyancy. As described by \citet[\emph{J. Fluid Mech.}  vol. 37,][pp. 51-80]{BaiWjfm1969a}, upon penetrating the interface the flow in each plume becomes forced and behaves like a constant-momentum jet, due to a reduction in its mean buoyancy relative to the local environment. To observe the behaviour of the plumes we partition the domain into sub-domains corresponding to each plume. Domains of relatively small aspect ratio produce a single primary mean-flow circulation between the sub-domains that is maintained by entrainment into the plumes. At larger aspect ratios the mean flow between the sub-domains bifurcates, indicating the existence of a secondary circulation within each layer associated with entrainment into the jets. The largest aspect ratios studied here exhibit an additional, tertiary, circulation in the vicinity of the interface. Consistency between independent calculations of an effective entrainment coefficient allows us to identify aspect ratios for which the flow can be modelled using plume theory, under the assumption of a two-layer stratification.

  To study the flow's energetics we use a local definition of available potential energy (APE). For plumes with Gaussian velocity and buoyancy profiles, the theory we develop suggests that the kinetic energy dissipation is split equally between the jets and the plumes and, collectively, accounts for almost half of the input of APE at the boundaries. In contrast, $1/4$ of the APE dissipation and background potential energy (BPE) production occurs in the jets, with the remaining $3/4$ occurring in the plumes. These bulk theoretical predictions agree with observations of BPE production from simulations to within $1\%$ and form the basis of a similarity solution that models the vertical dependence of APE dissipation and BPE production. Unlike results concerning the dissipation of buoyancy variance and the strength of the circulations described above, the model for the flow's energetics does not involve an entrainment coefficient.
  
\end{abstract}


\section{Introduction}

The turbulent plume is a canonical buoyancy-driven flow that plays
a fundamental role in a wide range of applications. Plumes induce
flow \citep{TayGmis1958a} and produce a buoyancy structure
\citep{WorMjfm1983a} in their surrounding environments, which in
turn affects their behaviour. This coupling is of particular
significance for applications involving confined spaces, such as
the heating and ventilation of individual rooms in a building
\citep{LinPafm1999a}, which are typically described using filling
box \citep{WorMjfm1983a} and emptying filling box
\citep{LinPjfm1990a} models.

Current understanding of how a plume is affected by confinement is
limited. Key questions in this regard relate to the effect of
background turbulence, stratification, mean co- or cross-flow and
entrainment or detrainment from nearby plumes \citep[see
e.g.][]{LaiAjfm2019a, KhoBjfm2013a,GlaCjfm2014a,BonRefm2018a}.
All of these effects have the potential to influence the buoyancy
structure of a space and the transport of pollutants. A particular
challenge to research is that such effects typically coexist and
each can influence the physics of plumes in several distinct
ways. This makes it difficult to develop models using reductionism
and might preclude simple explanations of observed phenomena. A
case in point that has received attention recently is the
observation and prediction that background turbulence reduces
entrainment into turbulent jets \citep{HunJjfm2006a, KhoBjfm2013a,
  LaiAjfm2019a}, in spite of contrary suggestions elsewhere
\citep{HubJphd2006a}. A further example from the present study is
our identification of a hierarchy of mean-flow structures
\citep[see][figure 10]{BaiWjfm1969a} induced by confined plumes,
which are necessarily accompanied by background turbulence and a
stable, layered stratification.

A central question concerning the various effects of confinement
is their influence on entrainment, which is the closure for
turbulence that underpins plume theory \citep{BatGqms1954a,
  MorBprs1956a}. The entrainment coefficient determines the
predicted ventilation and temperature of buildings
\citep{LinPafm1999a}, in addition to the spreading rate of plumes
\citep{MorBprs1956a}. Recent work has therefore sought to
establish a more precise and comprehensive understanding of the
wealth of physics for which the entrainment coefficient implicitly
accounts. The perspectives adopted range from studies of the
turbulent/non-turbulent interface at the local microscale
\citep{ReeMjfm2014a, SilCafm2014a, BurHjfm2017a} to consistent
relations with budgets for kinetic energy \citep{PriCqms1955a,
  KamEjfm2005a, ReeMjfm2015a} and buoyancy variance
\citep{CraJjfm2017a} from a global or integral perspective.

In spite of the leading role played by entrainment in relation to
force (buoyancy) and flow (velocity) in jets and plumes, and their
response to confinement, some results concerning energy budgets
(involving the combination of force and flow) do not require a
parameterisation of entrainment. An example is the stability
properties of integral models of unsteady jets and plumes
\citep{CraJjfm2015b, CraJjfm2016a}. A similar situation exists in
relation to the Nusselt number and the mechanical energy budgets
of Rayleigh-B\'{e}nard convection \citep{HugGjfm2013a}.  At high
Rayleigh numbers, the energy budgets for Rayleigh-B\'{e}nard
convection indicate that the input of available potential energy
(APE) is partitioned equally between viscous dissipation and the
energy required to homogenise the buoyancy field, which can be
stated as the `mixing efficiency' being equal to $1/2$. The
finding contrasts with results pertaining to the dissipation of
buoyancy variance \emph{per se} \citep[see e.g.][]{GroSjfm2000a},
which depends on the Nusselt number and does not play a direct
role in the system's energetics \citep{HugGjfm2013a}. An identical
result concerning the mixing efficiency of $1/2$ holds for a
filling box regardless of the value of the entrainment coefficient
\citep{WykMarx2018a}. In contrast, the expression for the mixing
efficiency of an \emph{emptying} filling box depends, in a
mathematical sense, on an entrainment coefficient
\citep{WykMarx2018a}.

Following \citet{PriCqms1955a}, a connection between entrainment
and a flow's budget for kinetic energy, the production of
turbulence kinetic energy and, therefore, viscous dissipation is
known \citep{KamEjfm2005a, ReeMjfm2015a}. Similar results,
concerning the dissipation of buoyancy variance in plumes, are
also known \citep{CraJjfm2017a}. However, as indicated in the
preceding paragraph, buoyancy variance \emph{per se} is not a
quantity that plays a direct role in a system's energetics. To
endow buoyancy variance with energetic implications, a system's
APE must be considered \citep{WinKjfm1995a}, which necessarily
introduces a mathematical dependence of the corresponding
dissipative quantity on the global probability distribution for
buoyancy.

Developments subsequent to \citet{WinKjfm1995a} that define local
budgets of APE \citep{ScoAjfm2006a, RouGjfm2009a, ScoAjfm2014a,
  NovLjas2018a}, following \citet{HolDjfm1981a} and
\citet{AndDjfm1981a}, provide the opportunity for establishing a
deeper understanding of energetics in the context of plume
modelling. However, local APE frameworks for diagnosing stratified
turbulence are still being developed and are relatively difficult
to apply. For example, recent contributions from
\citet{ScoAjfm2014a}, \citet{NovLjas2018a} and
\citet{TaiRjfm2018a} propose alternative ways to partition
mean and fluctuating components arising from turbulence.

An overarching aim of this work is to provide a bridge between the
classical description of problems involving plumes in terms of
entrainment and a deeper understanding of their energetics. In
addressing this aim, we will clarify the differences between those
aspects of confined convection that require a parameterisation for
entrainment and those that can be deduced from arguments
pertaining to the system's energetics alone. The latter are useful
because they provide constraints on the flow's energy
conversions. While we do not expect such constraints to restrict
the value of the entrainment coefficient, we expect the
information to be valuable more generally in the development and
understanding of bulk models for plumes in confined spaces. Our
hope is that the link we provide with the established global
energetics framework of \citet{WinKjfm1995a} will elicit
application of the local APE frameworks described in the previous
paragraph to diagnose aspects of the small-scale physics
associated with entrainment.

\begin{figure}
  \begin{center}
    \includegraphics[height=7.0cm,trim=70 610 70 50, clip]{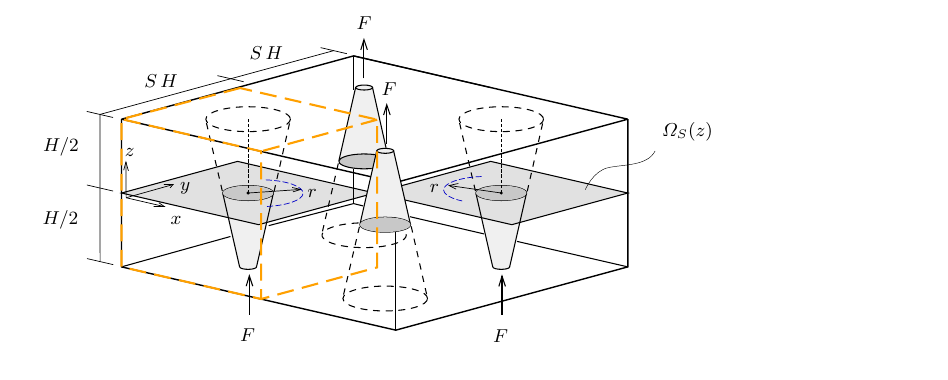}
\end{center}
\caption{Schematic arrangement of four plumes within a horizontally
  periodic domain. The horizontal shaded regions comprising
  $\Omega_{S}(z)$ highlight parts of the horizontal domain over which
  integrals are taken. The dashed quadrant corresponds to the
  sub-domain to which discussion in the main text refers, comprised
  of a turbulent plume in the lower layer and, nominally, a
  turbulent jet in the upper layer. The domain height is $H$, the quadrant aspect ratio is $S$, and the buoyancy flux at the plume sources is $F$.}
\label{diag:01}
\end{figure}
%

The flow we consider is driven by equal and opposite point sources
of buoyancy, as shown in figure \ref{diag:01}. The case provides a
convenient means of addressing several of the questions outlined
above, such as those relating to the interaction of the plumes
with each other and with a step change in ambient buoyancy
\citep{BaiWjfm1969a, CamRjfm2016a}. Indeed, the case is
complementary or dual to the approximately uniform mean buoyancy
produced in Rayleigh-B\'{e}nard convection, and therefore provides
a useful setting to investigate the flow's energetics in relation
to previous work \citep{HugGjfm2013a, WykMarx2018a}.

After describing the problem and simulation details in
\S\ref{sec:def}, we structure the paper as two parts. The first
(\S\ref{sec:mean}) deals with the buoyancy and flow structure,
which are aspects of the problem that relate explicitly to
entrainment. The second (\S\ref{sec:flow_energetics}), in
contrast, demonstrates that the flow's energetics can be
understood without reference to entrainment. 
\section{Problem definition}
\label{sec:def}
\subsection{Overview}

The case that we examine consists of an array of four plumes in a
horizontally periodic and square domain of height $H$ and
horizontal dimensions $2SH\times 2SH$, as shown in figure
\ref{diag:01}. Each diagonal pair of plumes is driven by sources
of buoyancy on either the top or the bottom of the domain. The
sources are located in the centre of each quadrant and each
provides a positive buoyancy flux $F$. For sufficiently large $S$,
the resulting steady state consists of a stable two-layer
stratification with a step change in buoyancy that is determined
by entrainment into the plumes. An equivalent case was described
in terms of line and point sources by
\citet[][p.72]{BaiWjfm1969a}.

\begin{figure}
  \begin{center}
  \begin{picture}(400,200)(0,-15)
    \put( 20,  0){\includegraphics[scale=0.23]{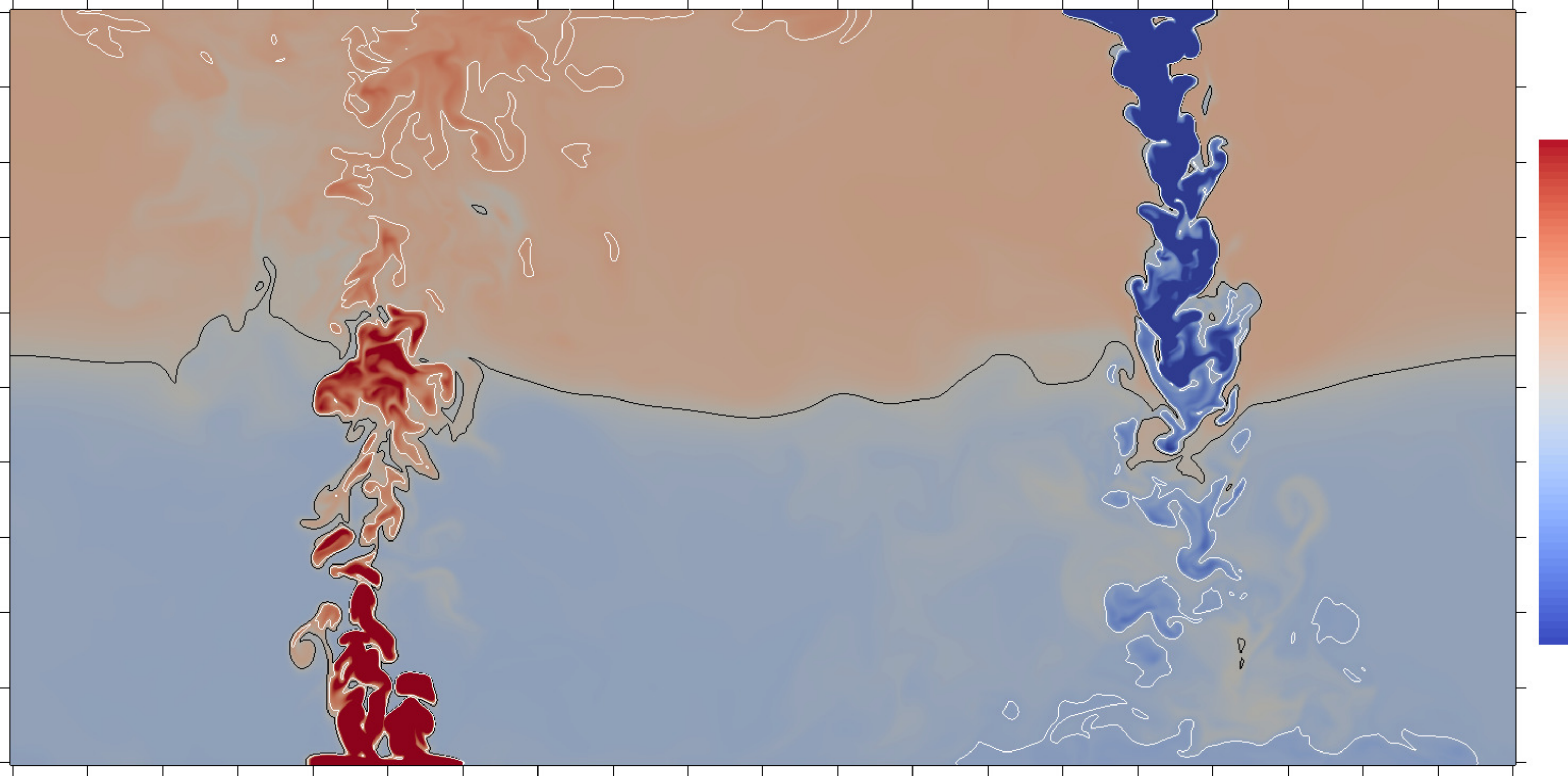}}
    \put( 0,  0){{\footnotesize $-0.5$}}
    \put( 5,  83){{\footnotesize$0.0$}}
    \put( 5,  165){{\footnotesize$0.5$}}
    \put( -5,  90){$z$}
    \put( 20,  -8){{\footnotesize $0.0$}}
    \put( 101,  -8){{\footnotesize $0.5$}}
    \put( 182,  -8){{\footnotesize $1.0$}}
    \put( 264,  -8){{\footnotesize $1.5$}}
    \put( 346,  -8){{\footnotesize $2.0$}}
    \put( 355,  19){$-20$}
    \put( 355, 142){$+20$}
    \put( 192,  -15){$x$}
  \end{picture}
  \end{center}
  \caption{The buoyancy field $b(x,z)$ over a vertical slice of
    the domain. The domain shown has a quadrant aspect ratio of
    $S=1$ and the slice intersects the vertical axis of two of the
    plumes. The black line denotes a buoyancy isosurface of zero and the
    two white lines denote buoyancy isosurfaces at $\pm 5.6$.}
  \label{fig:buoy}
\end{figure}
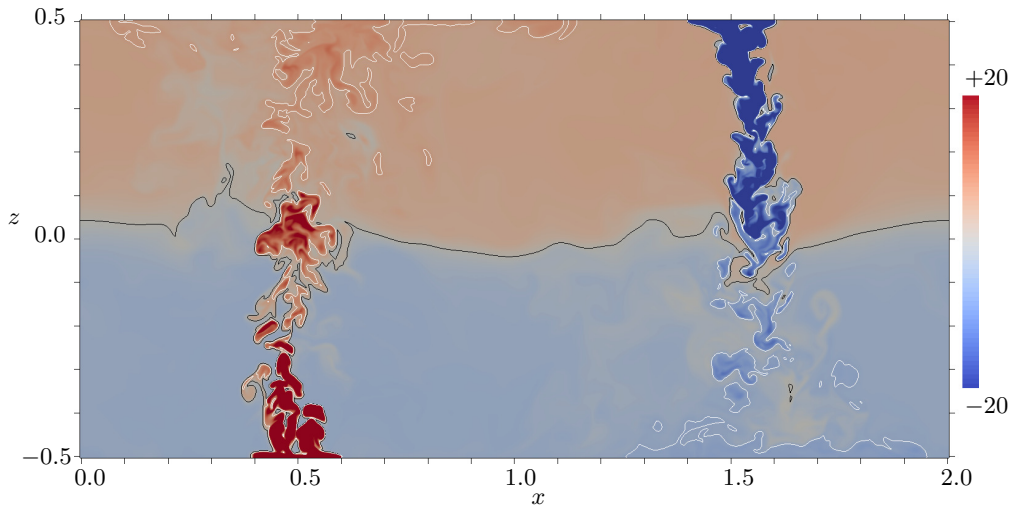

Figure \ref{fig:buoy} displays a vertical slice through the
instantaneous buoyancy field for $S=1$. The buoyancy structure
produced by localised sources of buoyancy contrasts with the
approximately uniform state that would be produced by the uniform
heating and cooling of the horizontal boundaries for Rayleigh
B\'{e}nard convection. In the lower layer above a heat source a
turbulent plume develops, whose fluid penetrates the interface
between the layers, before being entrained by the adjacent plumes.

\def\maxb{30}

\subsection{Governing equations}
\label{sec:label}

The equations of motion for the Boussinesq flow considered here
are

\begin{align}
  \pd{\vc{u}}{t}+\vc{u}\cdot\nabla\vc{u} &= -\nabla p + b\,\ez
+ \frac{1}{Re}\nabla^{2} \vc{u},
\label{eq:lmom}\\
\nabla\cdot \vc{u} &= 0, \label{eq:ldiv}\\
  \pd{b}{t}+\vc{u}\cdot\nabla b &= \frac{1}{Pe}\nabla^{2} b,\label{eq:lbuoy}
\end{align}

\noindent in which the scales used to non-dimensionalise velocity
$\vc{u}$, buoyancy $b$, kinematic pressure $p$, space $(x,y,z)$
and time $t$ are $F^{1/3}H^{-1/3}$, $F^{2/3}H^{-5/3}$,
$F^{2/3}H^{-2/3}$, $H$ and $F^{-1/3}H^{4/3}$, respectively. The
unit vector $\ez$, points in the vertical direction. The Reynolds
and P\'{e}clet numbers are $Re=F^{1/3}H^{2/3}/\nu$ and
$Pe=F^{1/3}H^{2/3}/\kappa$, respectively, for kinematic viscosity
$\nu$ and thermal diffusivity $\kappa$. In terms of dimensionless
variables, the buoyancy flux supplied by each source is equal to
unity. To estimate an integral time scale for this problem we
divide the quadrant volume $H^{3}S^{2}$ by the volume flux
$F^{1/3}H^{5/3}$, which results in a horizontal turnover time of
$F^{-1/3}H^{4/3}S^{2}$.

\subsection{Simulation details}

We will compare theoretical predictions to the direct simulation
of \eqref{eq:lmom}-\eqref{eq:lbuoy}. The code we use employs a
fourth-order finite volume discretisation, the details of which
are documented in \citet{CraJjfm2015a}. We simulate the medium of
air, with a Prandtl number $Pr=\nu/\kappa=0.71$ and consider flows with
Reynolds number $Re=4185$, which corresponds to a Rayleigh number
of $Pr\,Re^2 =1.24\times 10^{7}$. On the horizontal (top and
bottom) boundaries we apply a free-slip condition on the velocity
and homogeneous Neumann conditions on the buoyancy field outside
the circular sources. Inside the circular sources we impose an
inhomogeneous Neumann condition on buoyancy to produce a constant
flux. On the vertical boundaries (sides) of the domain we impose
periodic boundary conditions on all dependent variables. The grid
is Cartesian and uniformly spaced; hence the shape of the
nominally circular sources is formed approximately using square
cells to fill a disk of diameter $D=H/5$. The plumes are unforced
and therefore nominally lazy at their source
\citep{HunGjfm2005b}. This means that the plumes contract in the
vicinity of the source \citep{FanTjfm2003a}, resulting in an
effective radius that is significantly less than the physical
source radius (see, e.g. the sources in figure \ref{fig:buoy}). A
summary of the simulation details is provided in table
\ref{tab:02} and in appendix \ref{sec:valid} we demonstrate the
convergence of the results and investigate their sensitivity with
respect to changes in the Reynolds number.

\begin{table}
  \begin{center}
\def~{\hphantom{0}}
  \begin{tabular}{c@{\quad}r@{\quad}r}
    \vspace{2mm}
& $Nx\times Ny \times Nz$ & $S$\phantom{space}  \\
$\bigstar$ & $768^{2} \times 768$ &  1/2 = 0.50  \\
$\pentagon$ & $896^{2} \times 768$ & 7/12 = 0.58  \\
$\blacksquare$ & $960^{2} \times 768$ & 15/24 = 0.63  \\
$\Diamond$ & $1024^{2} \times 768$ & 2/3 = 0.67 \\
$\vartriangle$ & $1536^{2} \times 768$ & 1 = 1.00 \\
$\CIRCLE$ & $2048^{2} \times 768$ & 4/3 = 1.33 \\
  \end{tabular}
  \caption{Simulation details. All simulations were run at
    $Re = 4185$, $Pe = 2971$, and $DN_{x}/{2SH} = 154$, where
    $DN_{x}/{2SH}$ is the number of cells that span the source
    diameter $D$ for quadrant aspect ratio $S$. The symbols in the
    leftmost column correspond to those used in figures
    hereafter.}
  \label{tab:02}
  \end{center}
\end{table}

\subsection{Domain decomposition}

We take statistics over quadrants that are centred on each source,
as illustrated in figure \ref{diag:01}. Unlike the statistics that
can be obtained over the entire horizontal plane, the quadrant
statistics are, in general, vertically asymmetric with respect to
the interface and allow us to isolate the vertical evolution of
each plume. In addition, it proves convenient to respect the
symmetry of the problem by aligning the origin of the vertical
coordinate $z$ with the interface, as shown in figure
\ref{diag:01}.

We define time-averaged integrals by integrating over diagonal
pairs of quadrants $\Omega_{S}(z)$, for which the flow is
statistically equivalent (see, for example, the square shaded regions in
figure \ref{diag:01}), such that 

\begin{equation}
  \hav{f} (z) \equiv \frac{1}{2T}\int_{0}^{T}\iint_{\Omega_{S}(z)}f(\vc{x},t)\rd x\,\rd y\,\rd t, \label{eq:average}
\end{equation}

\noindent The time $T$ over which the integrals were averaged was
not less than $1$ dimensionless horizontal turnover time
$F^{1/3}H^{-4/3}S^{2}$. Whilst the factor of $1/2$ in
(\ref{eq:average}) means that $\hav{f}$ can be interpreted as an
integral over a `single' quadrant, we emphasise that
\eqref{eq:average} integrates information from a diagonal pair of
quadrants, in which the flow has the same sense, as shown in
figure \ref{diag:01}.

We will use double angles to denote integrals over a \emph{volume}
of the domain,

\begin{equation}
  \vav{ f }_{a}^{b}\equiv\int_{a}^{b}\hav{f}(z)\rd z.
\end{equation}

\noindent Acknowledging that our use of $\hav{}$ and $\vav{}$ to
denote \emph{integrals} rather than \emph{averages} is not
standard, we note that for this problem, in which we vary the
aspect ratio, it proves convenient. For example, the constant
area-integrated buoyancy flux at each plume source is independent
of a domain's aspect ratio, in contrast to the spatially averaged
buoyancy flux, which is inversely proportional to the aspect ratio
squared.

\section{Mean flow and buoyancy structure}
\label{sec:mean}

In this section we discuss the flow's mean velocity and buoyancy
structure in relation to the effects of entrainment and observe
their dependence on the domains' aspect ratio.

\subsection{Observations}
\label{sec:flow}

Convection above or below each source in figure \ref{fig:buoy}
results in turbulent plumes that entrain, dilute and transport
fluid between the layers and, in doing so, determine an
approximately two-layer stratification of the surrounding
environment \citep{BaiWjfm1969a}.  Buoyancy conservation indicates
that in a statistically steady state, in which the ambient
buoyancy in each layer does not change, the mean buoyancy in a
plume at the level of the interface is approximately equal to the
ambient buoyancy of the layer downstream. We will therefore refer
to the resulting neutrally buoyant flow into the downstream layer
as a turbulent jet (as can be seen in the top left and bottom
right of figure \ref{fig:buoy}). For discussion purposes we will
focus on the quadrant containing a plume oriented in the positive
vertical direction, as highlighted in figure \ref{diag:01}, and
will therefore refer to the plume as occupying the lower layer and
the jet as occupying the upper layer. In spite of the terminology
we employ, we can see in figure \ref{diag:01} that density
anomalies persist within the jets, rendering their
flow different from conventional uniform-density jets.

\begin{figure}
  \begin{center}
    \includegraphics[scale=1]{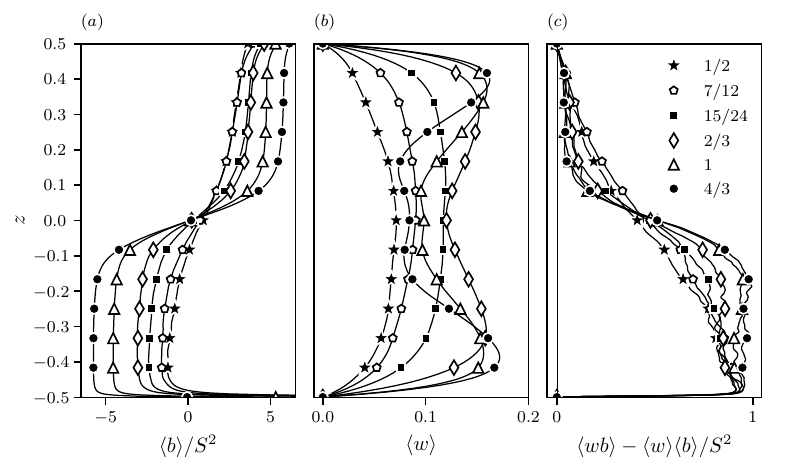}
  \end{center}
  \caption{Quadrant integrals of the average $(a)$ buoyancy
    (divided by $S^{2}$), $(b)$ volume flux, and $(c)$ relative
    buoyancy flux, for aspect ratios
    $S=1/2, \, 7/12, \, 15/24, \, 2/3, \, 1$, and $4/3$, as
    indicated by the symbols. The plume occupies the lower half of
    the domain, as highlighted in figure \ref{diag:01}.}
  \label{fig:01}
\end{figure}

Figure \ref{fig:01} indicates how quadrant integrals of buoyancy,
volume flux and relative buoyancy flux vary in the vertical
direction. Figure \ref{fig:01}$(a)$ shows that as the aspect ratio
of the domain increases, the difference between the buoyancy of
the lower and upper layer increases. The buoyancy profiles in
figure \ref{fig:01}$(a)$ are slightly asymmetric due to the
influence of the plume, whose mean buoyancy is non-zero, in the
lower layer. The asymmetry is relatively weak because the volume
occupied by the plume is small in comparison with that of the
domain.

The vertical volume flux is zero when integrated over the entire
domain, but non-zero when integrated over a single quadrant. By
examining the behaviour of the vertical volume flux in a single
quadrant, we can therefore identify the strength and structure of
the large-scale circulation. For example, if the vertical volume
flux $\hav{w}$ is increasing with height $z$ then, by continuity,
fluid is being drawn in through the sides of the quadrant
at that height $z$. If, on the other hand, the vertical volume
flux is decreasing with height, then there is a net flow of fluid
out of the quadrant through its sides.

\begin{figure}
  \begin{center}
    \includegraphics[scale=1]{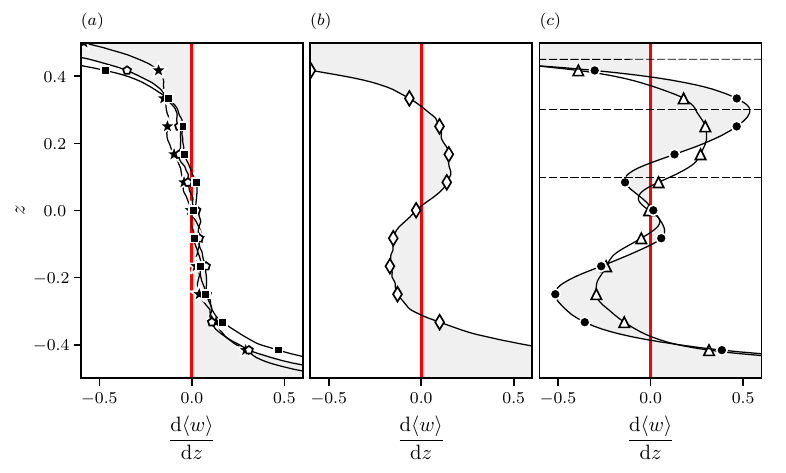}
  \end{center}
  \caption{The vertical derivative of the quadrant volume flux
    $\hav{w}$, which is equal to the horizontal flow through the
    vertical sides of the quadrant. The flow on domains of aspect
    ratio $S=1/2, 7/12, 15/24$ are shown in $(a)$ and do not
    contain any critical points. The flow on a domain of aspect
    ratio $S=2/3$ is shown in $(b)$ and contains two critical
    points and therefore bidirectional flow through the vertical
    sides of the quadrant in the upper and lower half of the
    domain. Flows on domains of aspect ratio $S=1, 4/3$ are
    shown in $(c)$ and contain four critical points, indicating
    the presence of a tertiary circulation cell at the
    interface. Horizontal cross-sections of the velocity field are
    plotted for the $S = 4/3$ case in figure \ref{fig:15}, at the
    heights indicated by dashed lines on $(c)$.}
  \label{fig:21}
\end{figure}

The vertical volume flux for a single quadrant is plotted in
figure \ref{fig:01}$(b)$. For the three smallest aspect ratios
$S=1/2, 7/12, 15/24$, figure \ref{fig:01}$(b)$ shows that the
volume flux increases in a quadrant above the source (located at
$z=-0.5$ in figure \ref{fig:01}), before decreasing in the upper
layer. This indicates that fluid is drawn into the quadrant
through the sides in the lower layer, transported vertically
between the layers and subsequently transported out of the
quadrant through the sides in the upper layer, as evidenced
explicitly in figure \ref{fig:21}$(a)$, which displays the
vertical derivative of the volume flux. From the perspective of
the spatially averaged flow, the entrainment and convection driven
by the plumes results in a single-cell large-scale circulation
between the quadrants, and volume conservation implies that these
processes are necessarily symmetric about the mid-plane $z=0$.

In a spatially averaged sense, a topological change in the flow
between quadrants occurs when the aspect ratio increases to
$2/3$. This can be identified from the emergence of two local
maxima in $\hav{w}$, as seen in figure \ref{fig:01}$(b)$, and the
bidirectional flow in the upper and lower halves of figure
\ref{fig:21}$(b)$. At a local maximum in $\hav{w}$, the mean
horizontal flow across the sides of the quadrant changes
direction, from a net flow into the quadrant to a net flow out of
the quadrant.  The appearance of two additional maxima therefore
indicates that secondary circulation cells have developed in each
layer. Noting that secondary circulations might exist
\emph{within}, rather than \emph{between}, each quadrant for
$S\leq 15/24$, we conclude that $S>15/24$ entails a secondary
circulation that engulfs more than half of a given layer.

The secondary circulation cells form due to entrainment into the
turbulent jets. When the plumes are sufficiently far apart, a given
jet entrains fluid in addition to that which is supplied by the
plume; hence the flux of volume that leaves the quadrant where the
jet impinges on the horizontal boundary of the domain exceeds the
total volume entrained by the plumes in the neighbouring quadrants
and the residual volume recirculates. The existence of secondary
circulation cells is therefore an indication that the plumes are,
to some extent, behaving independently. We will demonstrate in
\S\ref{sec:rprofs}, where we analyse the velocity profiles within
the jet, that the residual volume flux is indeed driven by
entrainment \emph{into} the jet, rather than being sustained as a
co-flow \emph{outside} the jet, which is not assured by the
information in figures \ref{fig:01} and \ref{fig:21}.

The additional volume flux due to entrainment into the jet can be
calculated using plume theory. We assume that the flow nominally
behaves as a turbulent jet above the interface. More precisely,
the flow is a forced plume, because figure \ref{fig:01}$(c)$
indicates a residual buoyancy flux in the flow above the
interface, which diminishes with aspect ratio. The volume flux in
the jet is equal to the volume flux in the plume at the interface
plus the volume flux due to the subsequent entrainment into the
jet,

\begin{equation}
Q(z) =  \underbrace{Q_{m}}_{\text{plume}} + \underbrace{\frac{6\alpha\sqrt{\pi}}{5}M_{m}^{1/2}z}_{\text{jet}},
\label{eq:Qjet}
\end{equation}

\noindent where $\alpha$ is the entrainment coefficient for a
plume and we have explicitly accounted for the fact that the
entrainment coefficient in plumes is approximately $5/3$ larger
than it is in jets \citep{ReeMjfm2015a}. Hereafter it proves
convenient to label the distance from the plume source to the
interface using the symbol $\zeta$.  Hence, the
quantities $Q_{m}$ and $M_{m}$ in \eqref{eq:Qjet}, corresponding
to the volume flux and momentum flux in the plume at the
interface, respectively are,

\refstepcounter{equation}
$$
Q_{m} = \frac{6\alpha}{5}\left(\frac{9\alpha\pi^{2}}{10}\right)^{1/3}\zeta^{5/3},\ \
M_{m} = \pi^{1/3}\left(\frac{9\alpha}{10}\right)^{2/3}\zeta^{4/3}.
\eqno{(\theequation{\mathit{a},\mathit{b}})}\label{eq:QM}
$$

\noindent Equations \specialeqref{eq:QM}{a} and
\specialeqref{eq:QM}{b} assume that the local (lower layer)
environment is unstratified; a stratification would result in
values of $Q_{m}$ and $M_{m}$ less than those predicted by
\specialeqref{eq:QM}{a} and \specialeqref{eq:QM}{b}. A comparison
of equation \eqref{eq:Qjet} at $z=0$ with equation
\specialeqref{eq:QM}{a} indicates that the rate at which the jet
entrains volume $Q(z=\zeta)-Q_{m}$ is \emph{equal} to the total
volume entrainment rate $Q_{m}$ of the plume. This property is a
consequence of the observed ratio of $5/3$ between the entrainment
coefficient in a plume compared with a jet
\citep{ReeMjfm2015a}. Consequently, for sufficiently large aspect
ratios, the strength of the secondary circulation is equal to the
strength of the primary circulation, as can be verified from
figure \ref{fig:01}$(b)$, which shows that the maximum value of
$\hav{w}$ is twice as large as its value at $z=0$ for $S=4/3$. The
maximum volume flux in a quadrant results from entrainment into
the plume in addition to fluid that will be re-entrained by the
adjacent jet.

\begin{figure}
  \begin{center}
    \includegraphics[scale=1]{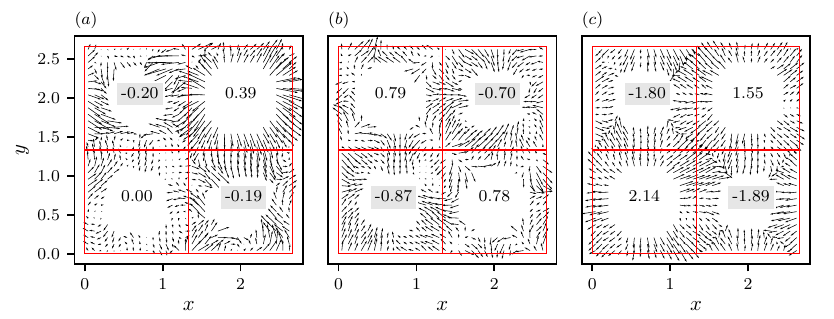}
  \end{center}
  \caption{Horizontal slices through the time-averaged velocity
    field and the horizontal divergence
    $\hav{\partial_{x}u+\partial_{y}v}$ (displayed in text) at
    $(a)$ $z = 0.1$, $(b)$ $z = 0.3$, $(c)$ $z = 0.45$ for aspect
    ratio $S=4/3$. The locations of these slices are plotted as
    dashed lines on figure \ref{fig:21}$(c)$. Note that persistent
    large scale structures in the horizontal flow result in
    time-averaged velocity fields in diagonal pairs of quadrants
    that are not necessarily identical.}
  \label{fig:15}
\end{figure}

For the domains of relatively large aspect ratio, such as $S=1$
and $S=4/3$, figure \ref{fig:01}$(b)$ shows that a further
bifurcation in the mean-flow structure occurs. The change
corresponds to the emergence of a third, relatively weak,
circulation cell, which can also be identified from the two
additional critical points in $\hav{w}$, as shown in figures
\ref{fig:01}$(b)$ and \ref{fig:21}$(c)$. The three circulation
cells for $S=4/3$ can also be observed in the horizontal velocity
fields shown in figure \ref{fig:15}. The quadrants for which the
integrals in figures \ref{fig:21} and \ref{fig:15} were calculated
correspond to the bottom left and top right quadrants. Just above
the interface, where $z = 0.1$, there is a net inflow into the top
left and bottom right quadrants and a net outflow from the top
right quadrant. At $z = 0.3$, the divergence is stronger and
indicates that the flow entering or leaving a given quadrant has
reversed in comparison with $z=0.1$. At $z=0.45$ the direction of
flow entering and leaving each quadrant changes for a second time
and corresponds to the primary circulation driven by entrainment
into the plumes.

We close our analysis of the mean flow's topology by noting a
likely influence of free-slip velocity boundary conditions in
determining the shape and extent of the circulation cells. Such
boundary conditions are convenient in allowing one to focus
exclusively on the flow that is naturally driven by the plumes, in
the absence of the competing, and aspect ratio dependent,
effects arising from wall friction. We recognise, however,
that practical problems involving confined turbulent plumes
necessarily involve wall friction, in addition to the phenomena on
which we focus.

For domains of sufficiently large aspect ratio, mass conservation
implies that the vertical flux of buoyancy, relative to the
ambient buoyancy, $\hav{wb}-\hav{w}\hav{b}/S^{2}$ within a quadrant will
be equal to unity in the layer containing the plume and zero in
the layer containing the jet. This argument requires the ambient
buoyancy in each layer to be uniform, such that the flux of
buoyancy, relative to the ambient is zero. Figure
\ref{fig:01}$(c)$ indicates that this is approximately correct for
the two largest aspect ratios $S=1, \, 4/3$, excepting a
relatively small residual flux in the upper layer. In the vicinity
of the top boundary at $z=1/2$, the relative buoyancy flux
$\hav{wb}-\hav{w}\hav{b}/S^{2}$ reduces to zero and the buoyancy in
figure \ref{fig:01}$(a)$ increases, which suggests that the
residual buoyancy is transported out of the quadrant horizontally
in a steady axisymmetric gravity current.

\subsection{Entrainment}
\label{sec:entrainment}

In general, flows driven by turbulent plumes are interpreted and
modelled using plume theory and an entrainment coefficient, under
the assumption that each plume acts independently. Entrainment can
be estimated from observed volume fluxes or buoyancy
differences. In this section we will calculate effective
entrainment coefficients $\alpha_{*}$ from observations of volume
flux and buoyancy, and use the estimates to identify aspect ratios
for which plume theory provides a faithful description of the
flow.

In the context of plume theory, the increase in the volume flux in
a plume is equal to the rate of entrainment from the surrounding
environment. Entrainment is typically parameterised as being
proportional to the square root of the momentum flux in the flow,
as discussed in \citet{BaiWjfm1969a}. The resulting coefficient of
proportionality for an isolated unconfined plume is
$\alpha\approx 0.12$ \citep[cf.][]{CarGjfm2006a, ReeMjfm2015a}.

Following \citet{BaiWjfm1969a}, there are two ways to calculate an
entrainment coefficient from the observations reported here. One
method is to find the entrainment coefficient from the volume flux
at the mid-plane. This could be considered the \emph{ab initio}
estimation of entrainment, because it describes the rate of
increase of the volume flux in the plume directly. Alternatively,
we can calculate the entrainment coefficient that would be
required to predict the observed difference in buoyancy between
the upper and lower layers under the assumption that mixing occurs
exclusively within the plumes. If the two entrainment rates are
equal, they suggest that plume theory provides a faithful
description of the flow.

For transparency, we refrain from using a virtual source
\citep{HunGjfm2001b} to account for the finite area of the sources
used in the simulations. Indeed, the theoretical location of the
virtual source of the infinitely lazy plumes (zero source momentum
flux) in this problem coincide with the vertical location of the
actual sources, i.e.\ the theoretical virtual source correction is
zero \citep{HunGjfm2001b}. However, figure \ref{fig:buoy} suggests
that, unlike the idealised contraction of lazy plumes that is
predicted by plume theory, the near-field behaviour of the plumes
in this study is disrupted by background turbulence; hence the use
of an effective virtual source might be appropriate. We therefore
note that by proceeding without using a virtual source, which
would increase $\zeta$ (the distance between the plume source and
the interface) in \eqref{eq:alphaQ} and \eqref{eq:alphab} below,
our estimations of the entrainment coefficient are likely to be
slightly larger than they would be if a virtual source were
incorporated.

\begin{figure}
  \begin{center}
    \includegraphics[scale=1]{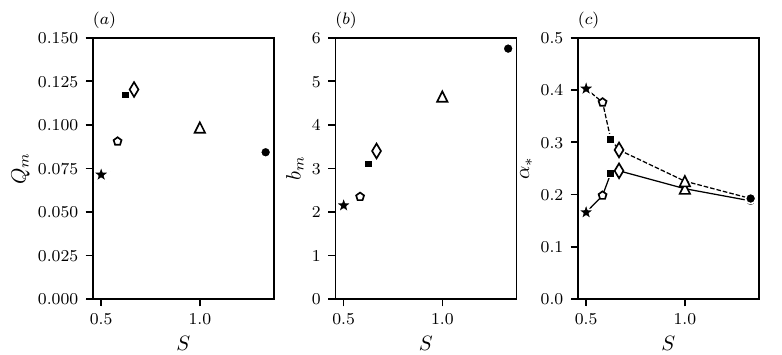}
  \end{center}
  \caption{An effective entrainment coefficient using plume theory:
    $(a)$ The volume flux at the interface $Q_m$, $(b)$ half the
    difference in buoyancy between the layers $b_m$, and $(c)$ the
    effective entrainment coefficient $\alpha_*$ for various
    aspect ratios $S$ (see table \ref{tab:02}).  An estimate for the entrainment rate
    $\alpha_Q$ (solid line) can be calculated from $Q_m$ and
    equation (\ref{eq:alphaQ}). A second estimate for the
    entrainment coefficient $\alpha_b$ (dashed line) is the value
    required to explain the observed buoyancy difference $b_m$
    between the layers (\ref{eq:alphab}).}
  \label{fig:entrainment_rate}
\end{figure}

First we will calculate an effective entrainment coefficient from the volume flux
at the interface. Inverting \specialeqref{eq:QM}{a} provides a means 
of estimating the entrainment coefficient from observations of the 
volume flux at the interface,

\begin{equation}
  \alpha_Q = \left(\frac{5}{6}\right)^{3/4}\left(\frac{10}{9\pi^{2}}\right)^{1/4}\frac{Q_{m}^{3/4}}{\zeta^{5/4}}.
\label{eq:alphaQ}
\end{equation}
where $\zeta$ is the distance from the plume source to the interface and 
$Q_{m}$ is the volume flux at the level of the interface (plotted
in figure \ref{fig:entrainment_rate}$(a)$).
The entrainment coefficient estimated from (\ref{eq:alphaQ}) is plotted 
against the aspect ratio $S$ in figure \ref{fig:entrainment_rate}$(c)$ (solid line). 

From figure \ref{fig:entrainment_rate}$(c)$ we see that the
entrainment coefficient, $\alpha_Q$, calculated from the volume
flux at the interface $Q_m$, increases with aspect ratio $S$ up to
some critical aspect ratio of approximately $S \approx
2/3$. Interestingly, this is close to the aspect ratio for which
we observed a topological change in the mean flow between
quadrants and the emergence of the secondary circulation cells in
the upper and lower layers (cf. \S\ref{sec:flow}). As
\eqref{eq:alphaQ} assumes an unstratified ambient, the increase in
the estimate of $\alpha_Q$ with increasing aspect ratio $S$ at low
aspect ratios may be due to the reduction of the stratification in
the lower layer as the aspect ratio increases. Increasing the
aspect ratio further results in a reduction in the entrainment
coefficient, which appears to be tending towards a value that is
slightly higher than the value of $0.12$ that is commonly
associated with an isolated plume in an unconfined environment.

A second method for calculating an effective entrainment
coefficient is from the difference in buoyancy between the layers.
In a steady state, mass conservation implies that the buoyancy of
a given layer is approximately equal to the mean buoyancy
$\pm b_{m}$ of the plume supplying fluid to that layer. We define
the relative buoyancy flux in each plume as being equal to the
volume flux $Q$ multiplied by the buoyancy of the plume relative
to the buoyancy of the local environment. The buoyancy flux of a
plume just beneath the interface is therefore
$Q_{m}(b_m - (-b_m)) = 2Q_{m}b_m$. Assuming an unstratified
environment in each layer, the buoyancy flux in the plume must be
equal to the dimensionless source buoyancy flux, therefore
$2Q_{m}b_m = 1$, which implies that $b_m = 1/(2Q_{m})$.

Noting that $b_{m}=1/(2Q_{m})$ and considering
\specialeqref{eq:QM}{a}, the buoyancy of the upper layer should
obey
\begin{equation}
  b_{m} = \frac{5}{12\alpha}\left(\frac{10}{9\alpha\pi^{2}}\right)^{1/3}\frac{1}{\zeta^{5/3}}.
\label{eq:bm}
\end{equation}
Strictly, the relationship \eqref{eq:bm} between the buoyancy of a
layer and the mean buoyancy of the plume by which it is fed is
approximate because a small proportion of the plume's total
buoyancy flux is provided by turbulent transport, which is
formally independent of the mean buoyancy in the plume. In other
words, the actual relative buoyancy flux in the plume just beneath
the interface is $2Q_{m}b_{m}$ plus a contribution of
approximately $15\%$ from turbulent transport
\citep{ReeMprf2016a}. Consequently, the buoyancy of a given layer
is typically slightly greater in magnitude than the mean buoyancy
of the plume might suggest.

We determine $b_{m}$ as half the distance between
peaks in the time-averaged probability density function for
buoyancy, which is shown in figure \ref{fig:pdf} and will be
discussed in \S\ref{sec:energy}. The value of $b_{m}$ for each 
value of $S$ is plotted in figure \ref{fig:entrainment_rate}$(b)$.
Knowledge of $b_{m}$ and use of 
equation \eqref{eq:bm} means that the entrainment coefficient can 
be estimated independently of \eqref{eq:alphaQ} from
\begin{equation}
  \alpha_b = \left(\frac{5}{12}\right)^{3/4}\left(\frac{10}{9\pi^{2}}\right)^{1/4}\frac{1}{b_{m}^{3/4}\zeta^{5/4}}.
\label{eq:alphab}
\end{equation}
The entrainment coefficient as calculated from equation \eqref{eq:alphab}
is plotted in figure \ref{fig:entrainment_rate}$(c)$ (dashed line). 
This entrainment rate can be considered as the entrainment rate 
required as input into a plume model to explain the observed 
buoyancy difference between the layers.

We observe from figure \ref{fig:entrainment_rate}$(c)$, that the
entrainment coefficient calculated from $b_m$, using equation
\eqref{eq:alphab}, results in a value of $\alpha_b$ that is
everywhere greater than the value of $\alpha\approx 0.12$ that is
commonly assigned to isolated and unconfined plumes. Indeed, the
use of $\alpha\approx 0.12$ would significantly overestimate the
buoyancy difference between the upper and lower layers. Figure
\ref{fig:entrainment_rate}$(c)$ shows that calculating an
entrainment coefficient from the buoyancy field using equation
\eqref{eq:alphab}, results in a larger estimate for $\alpha$ than
equation \eqref{eq:alphaQ}, which is based on the volume flux. The
difference can be attributed to two distinct effects. First, the
estimation \eqref{eq:alphab} implicitly assumes that vertical
buoyancy transport can be assigned exclusively to the plumes,
rather than interfacial mixing. Secondly, \eqref{eq:alphaQ} does
not account for stratification within each layer, which is
significant for domains with relatively small aspect ratio and
would lead to an increase in the corresponding prediction of
$\alpha_b$ and a decrease in the predictions of $\alpha_Q$.

Figure \ref{fig:entrainment_rate} indicates that for $S = 4/3$
equations \eqref{eq:alphaQ} and \eqref{eq:alphab} provide almost
identical and therefore robust estimates for $\alpha$, which
suggests that the effects of interfacial mixing are insignificant
for $S\geq 4/3$. However, the estimated value
$\alpha_{*} \approx 0.2$ for $S = 4/3$ is larger than
$\alpha\approx 0.12$ for an isolated plume in an unconfined
domain. In this regard, the use of an effective virtual source, a
distance $\lambda\zeta$ further from the interface than the actual
source, would lead to an estimate of
$\alpha_{*}\approx 0.2/(1+\lambda)^{5/4}$. However, noting that
classical plume theory predicts the virtual source correction for
infinitely lazy plumes to be zero \citep{HunGjfm2001b}, we would
associate an effective virtual source with the \emph{indirect}
effect that background turbulence has in modifying the near-field
development of the plume. In view of the reduction in entrainment
with respect to background turbulence reported by
\citep{LaiAjfm2019a, KhoBjfm2013a}, attribution of
$\alpha_{*}=0.2$ solely to \emph{direct} effects of background
turbulence would therefore be incorrect and misleading, not least
because it would also ignore possible modifications to entrainment
arising from the mean ambient flow discussed in \S\ref{sec:mean}.

We conclude by noting that the quadrant aspect ratios $S=1,4/3$
approximately correspond to the critical aspect ratio $H/R=1$ for
overturning identified by \citet{BaiWjfm1969a}, if the radius $R$
of the cylinder containing their single plume is compared with the
shortest distance $SH$ between the plume axes of the present problem.

\section{Flow energetics}
\label{sec:flow_energetics}

Unlike the volume flux $Q_{m}$ and the upper and lower layer
buoyancy $\pm b_{m}$, there are quantities that can be predicted
directly from the system's energy budget. An example is the total
viscous dissipation, which is necessarily equal to the volumetric
integral of the buoyancy flux, regardless of the strength of the
circulations and buoyancy differences studied in the previous
section. In the following section, we utilise such properties to
derive models for the flow's energetics that do not involve
an entrainment coefficient.

We will start by describing the energetics framework, defining the
local available potential energy and background potential energy
(BPE) in \S\ref{sec:energy}. We then discuss the viscous
dissipation and BPE production in \S\ref{sec:energy_stats}, before
describing models in
\S\ref{sec:ke_model}-\S\ref{sec:similarity}. In \S\ref{sec:rprofs}
and \S\ref{sec:ape_transport} we examine the similarity arguments
upon which these models are based.

\subsection{Governing equations for energetics}
\label{sec:energy}

\begin{figure}
  \begin{center}
    \includegraphics[scale=1]{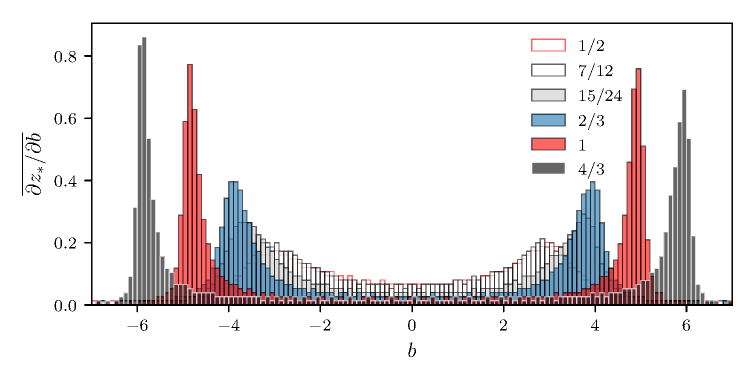}
  \end{center}
  \caption{The time-averaged probability density function of the
    buoyancy, which corresponds to $\partial z_{*}/\partial b$,
    over the entire domain for different aspect ratios.}
  \label{fig:pdf}
\end{figure}

Let $z_{*}(b,t)$ represent an adiabatic rearrangement of the fluid
into the configuration possessing the minimal potential energy
and, therefore, a monotonically increasing reference buoyancy
$b_{*}(z,t)$. The minimal potential energy is equal to the BPE
\citep{LorEtel1955a}. For this problem, in which
$z_{*}\in [-0.5,0.5]$, the quantity $\partial z_{*}/\partial b$
corresponds to the probability density function for
buoyancy. Indeed, $z_{*}$ can be obtained during a simulation by
integrating the probability density function for buoyancy
\citep{WinKjfm1995a}. The histograms of the time average of
$\partial z_{*}/\partial b$ for different aspect ratios shown in
figure \ref{fig:pdf} demonstrate that as the domain aspect ratio
increases the variance of the global distribution of buoyancy
increases.

When non-dimensionalised, the local gravitational potential energy
of the fluid is $E_{p}\equiv -bz$. Typically, the portion of the
potential energy that is available to do work to increase the
kinetic energy of the flow is computed as the volume integral of
$b(z_{*}-z)$ \citep{WinKjfm1995a}. Locally, however, $b(z_{*}-z)$
is not positive definite, which motivates the definition of a
suitable local APE density. Following
\citet{ HolDjfm1981a} and \citet{ScoAjfm2014a}, we therefore
consider the positive semidefinite quantity

\begin{equation}
  E_{a}\equiv \int_{b_{*}}^{b}(z_{*}(\hat{b},t)-z)\rd\hat{b}.
\label{eq:Ea}
\end{equation}

Physically, the definition \eqref{eq:Ea} is positive semidefinite
because it accounts for both the potential energy associated with
a parcel of fluid displaced from equilibrium and the potential
energy associated with the corresponding displacement of the
environment, as can be seen by decomposing \eqref{eq:Ea},

\begin{equation}
  E_{a}= b(z_{*}-z) - \int_{0}^{z-z_{*}}b_{*}(z-\hat{z},t)\rd\hat{z},
\label{eq:Ea2}
\end{equation}

\noindent where $b_{*}(z,t)$ is the reference buoyancy, such that
$z_{*}(b_{*}(z,t),t)=z$, corresponding to the adiabatically
rearranged state possessing minimal potential energy. The quantity
$E_a$ therefore has a dependence on the structure of the entire
buoyancy field, i.e.\ the value of $E_a$ at a point may change
because parcels of fluid mixing at locations that are remote from
that point leads to a modification of $b_*$. In the problem that
we consider the reference buoyancy can be equated with the ambient
buoyancy $\pm b_{m}$, except over thin layers at the bottom,
middle and top of the domain. Using \eqref{eq:lbuoy}, the
Lagrangian derivative of $E_{a}$ satisfies \citep{ScoAjfm2014a}

\begin{equation}
	\ld{E_a}{t} = \frac{\nabla^2 E_a}{Pe} + \frac{2}{Pe}\frac{\partial (b - b_*)}{\partial z} - G - w (b - b_*) + \int^b_{b_*} w_*(\hat{b},t) \rd \hat{b},
	\label{eq:ape_budget}
\end{equation}
\noindent where the APE dissipation 

\begin{equation}
G \equiv \frac{1}{Pe}\left(|\nabla b|^2 \left.\frac{\partial z_*}{\partial b}\right|_b - |\nabla b_*|^2 \left.\frac{\partial z_*}{\partial b}\right|_{b_*} \right),\quad\mathrm{and}\quad w_{*}\equiv \pd{z_{*}}{t}. \label{eq:G}
\end{equation}

The final term in (\ref{eq:ape_budget}) is zero when the
probability density function for buoyancy (cf. figure
\ref{fig:pdf}) does not depend on time, which is true in the
present context for domains that are of sufficiently large aspect
ratio. In that case, equation \eqref{eq:ape_budget} reduces to

\begin{equation}
  \pd{E_{a}}{t}+\nabla\cdot(\vc{u}E_{a})=\frac{\nabla^{2}E_{a}}{Pe}+\frac{2}{Pe}\pd{(b-b_{*})}{z}
  -G-\underbrace{w(b-b_{*})}_{-\Phi_{z}},
  \label{eq:ape}
\end{equation}

\noindent in which $-\Phi_{z}$ is a buoyancy flux relative to the
reference state.

Following \citet{ScoAjfm2014a}, we define the local BPE density
$E_{b}$ as $E_{b}\equiv -bz-E_{a}$, whose budget obeys

\begin{equation}
  \pd{E_{b}}{t}+\nabla\cdot(\vc{u}E_{b})=\frac{\nabla^{2}\psi}{Pe}-\nabla\cdot(\vc{u}p_{*})
  +\underbrace{\left.\frac{ |\nabla b|^{2}}{Pe}\pd{z_{*}}{b}\right|_{b}}_{\Phi_{d}},
  \label{eq:bpe}
\end{equation}
\noindent where
\begin{equation}
  \psi = -\int_{0}^{b}z_{*}(\hat{b})\rd \hat{b},\quad\quad
  p_{*} = \int_{0}^{z}b_{*}(\hat{z})\rd \hat{z}.
\end{equation}
\noindent In a Boussinesq flow, the mixing rate $\Phi_{d}$
represents an irreversible conversion of APE into BPE via
diapycnal mixing. An interesting feature of $\Phi_{d}$, which
proves to be crucial when considering energetics and entrainment,
is that it corresponds to the dissipation of buoyancy variance
$|\nabla b|^{2}/Pe$ weighted by the probability density function
$\partial_{b}z_{*}$. Diapycnal mixing therefore has more energetic
significance when it accounts for mixing in regions of buoyancy
whose probability density is relatively large.

In the subsequent analysis, we choose to focus on the production
of BPE $\Phi_{d}$ rather than the APE dissipation $G$, because it
is the volume integral of $\Phi_{d}$ that features explicitly in
previous work on bulk energetics \citep[][for
example]{WinKjfm1995a, HugGjfm2013a} and evaluates to unity in
this particular problem (see equation \eqref{eq:phi_d_unity}). We
note that locally BPE production, $\Phi_{d}$, is not necessarily
equal to APE dissipation $G$, because the latter only accounts for
mixing resulting from macroscopic motion in the flow and not the
mixing $(\partial z_{*}/\partial b)|\nabla b_{*}|^{2}/Pe$
associated with diffusion of the background state; hence
$\Phi_{d} \geq G$ for the Boussinesq model considered
here. However, in the present problem, the rearranged state
consists primarily of two constant density layers in which
$\nabla b_* \approx 0$ (see, for example, figure
\ref{fig:pdf}). With the possible exception of thin layers near
the horizontal boundaries, $\Phi_{d}$ and $G$ are therefore
approximately equal, because they are dominated by the large
gradients of $b$ within the plumes. In regarding local BPE
production as equivalent to APE dissipation for this particular
\emph{Boussinesq} flow we are not endorsing the use of BPE
production to characterise mixing more generally. The energy
conversion implied by BPE production is different to APE
dissipation \citep{TaiRjfm2009a} and in \emph{non-Boussinesq}
models, for which BPE `production' can be negative
\citep{TaiRjfm2009a, GreMams2018a}, the difference is significant.

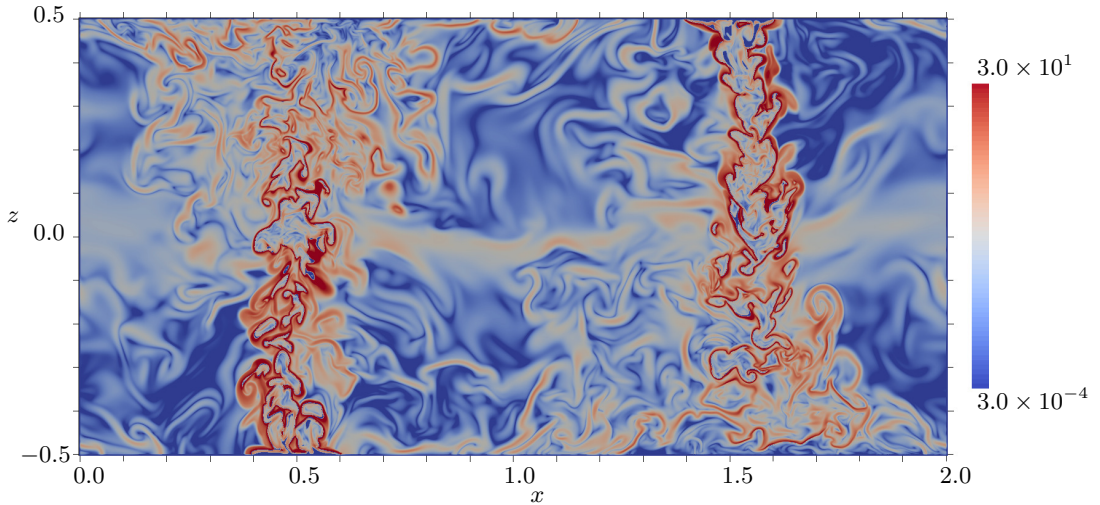
\begin{figure}
  \begin{center}
  \begin{picture}(400,200)(0,-15)
    \put( 20,  0){\includegraphics[scale=0.23]{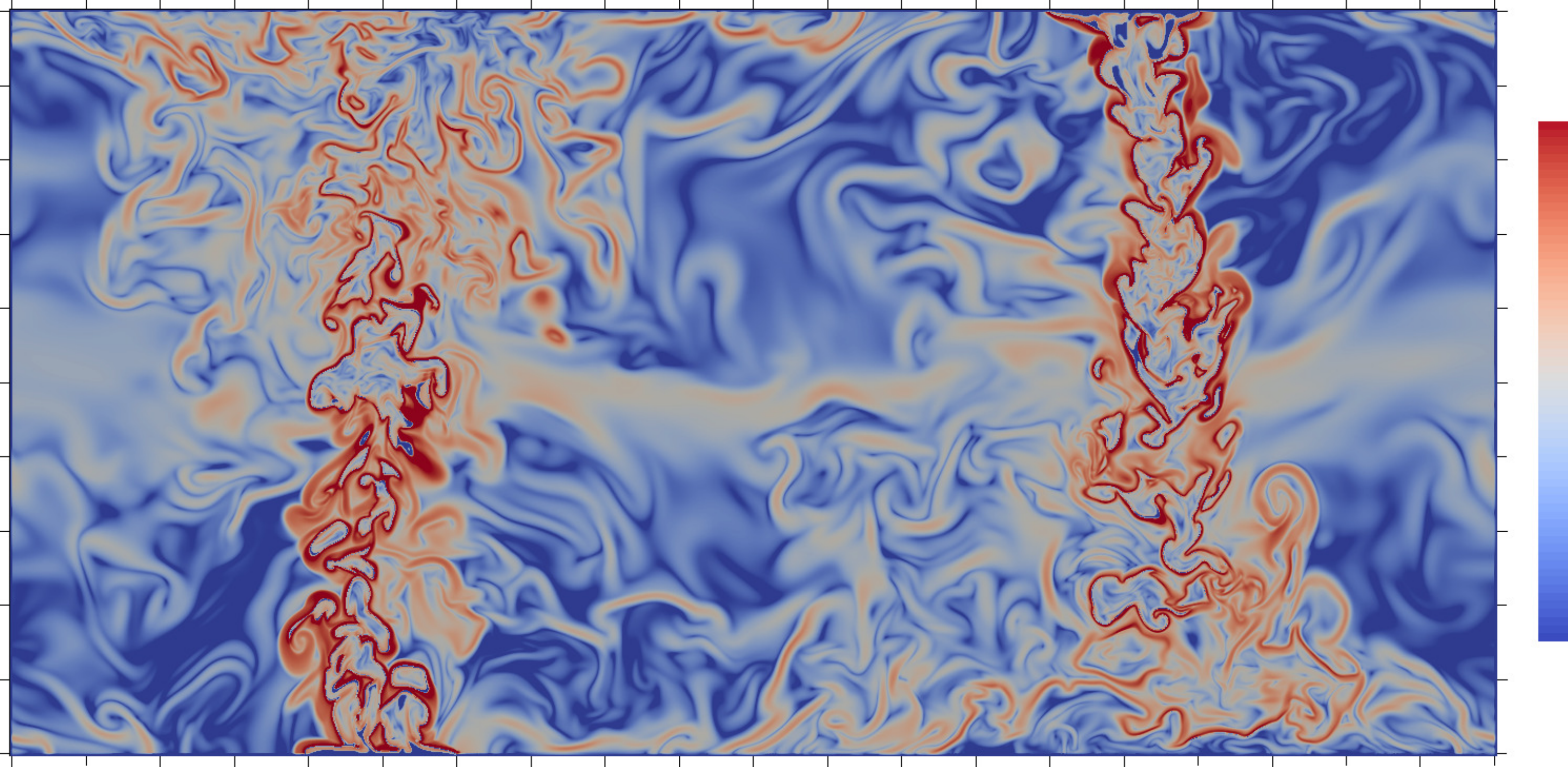}}
    \put( 0,  0){{\footnotesize $-0.5$}}
    \put( 5,  83){{\footnotesize$0.0$}}
    \put( 5,  165){{\footnotesize$0.5$}}
    \put( -5,  90){$z$}
    \put( 20,  -8){{\footnotesize $0.0$}}
    \put( 101,  -8){{\footnotesize $0.5$}}
    \put( 182,  -8){{\footnotesize $1.0$}}
    \put( 264,  -8){{\footnotesize $1.5$}}
    \put( 346,  -8){{\footnotesize $2.0$}}
    \put( 360,  20){$3.0\times 10^{-4}$}
    \put( 360, 145){$3.0\times 10^{1}$}
    \put( 192,  -15){$x$}
  \end{picture}
  \end{center}
  \caption{The BPE production
    $\Phi_{d}=(\rd z_{*}/\rd b)|\nabla\,b|^{2}/Pe$ over a vertical slice of
    the domain using a logarithmic scale. The domain has a
    quadrant aspect ratio of $S=1$ and the slice intersects the
    vertical axis of two of the plumes. The slice was taken at the
    same time as that of the buoyancy field displayed in figure
    \ref{fig:buoy}.}
  \label{fig:vslices}
\end{figure}

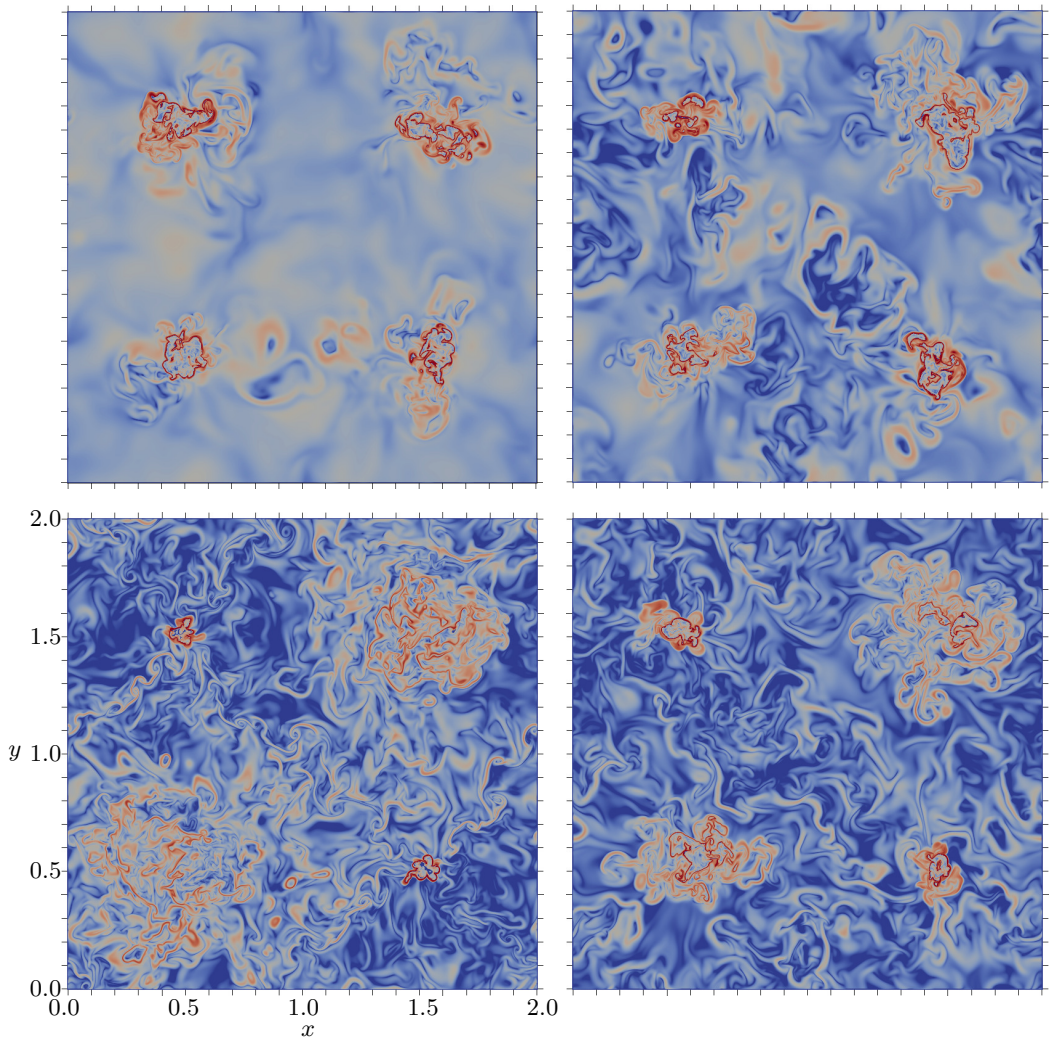
\begin{figure}
  \begin{center}
  \begin{picture}(400,400)(0,-15)
    \put( 10,  190){\includegraphics[scale=0.23]{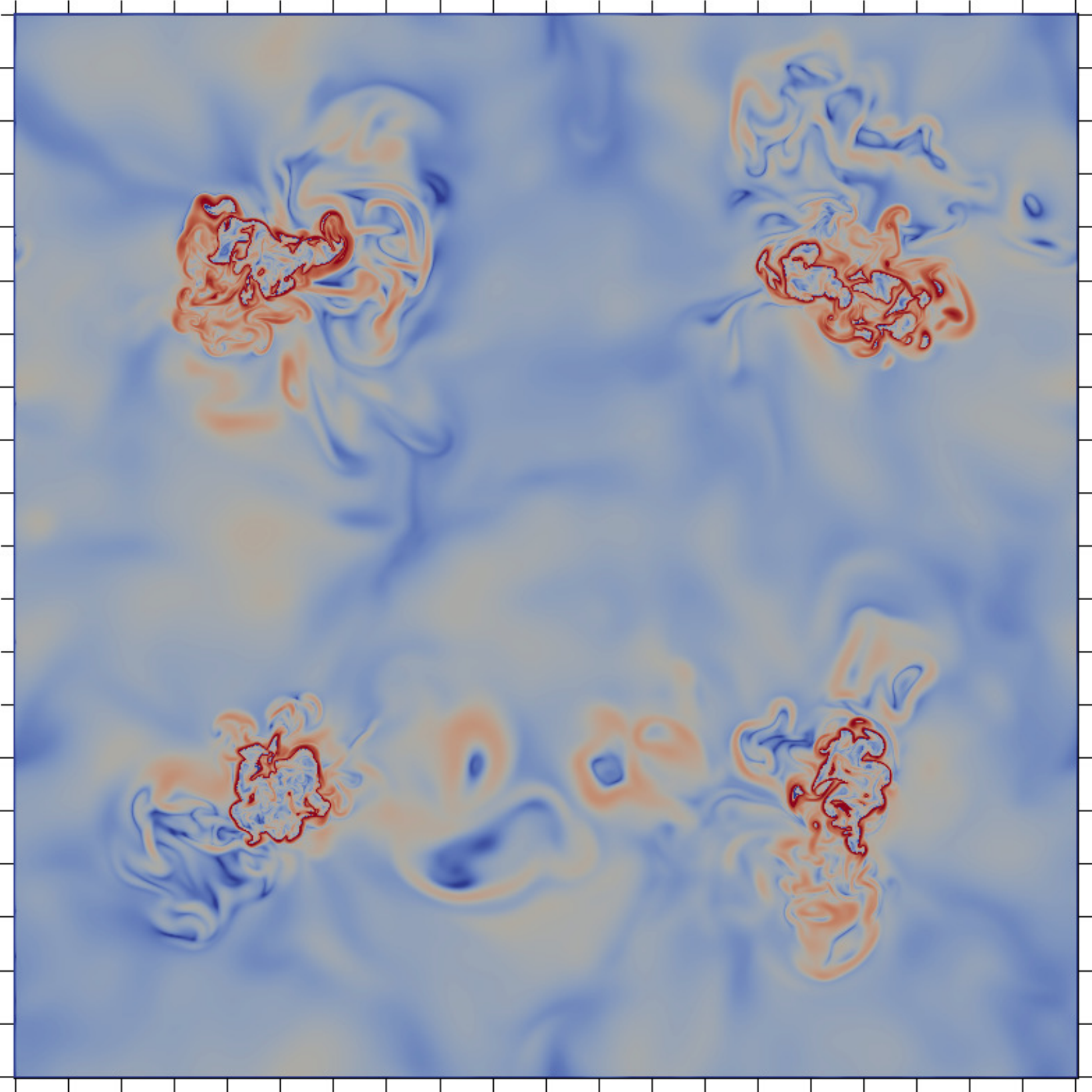}}
    \put( 200, 190){\includegraphics[scale=0.23]{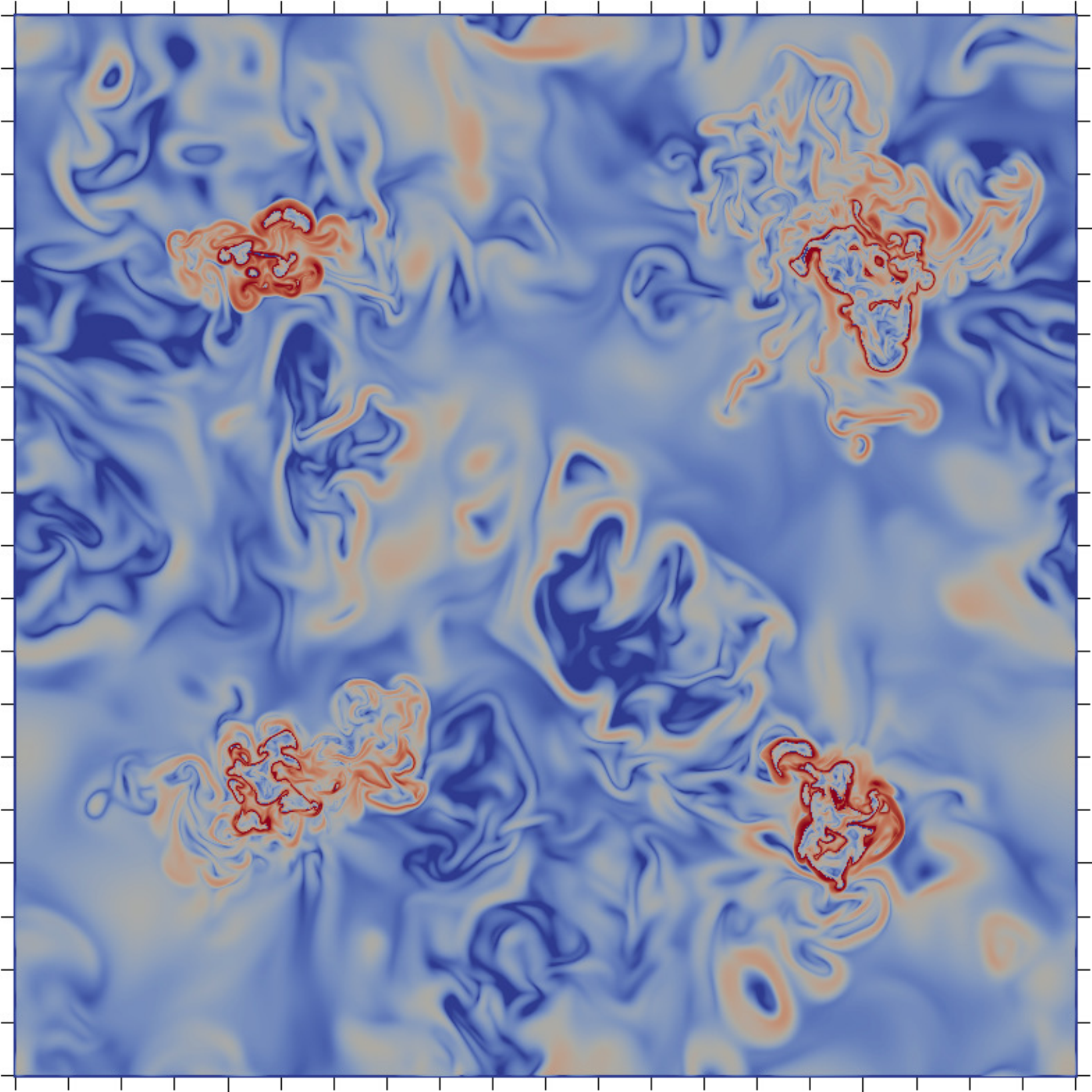}}
    \put( 200,  0){\includegraphics[scale=0.23]{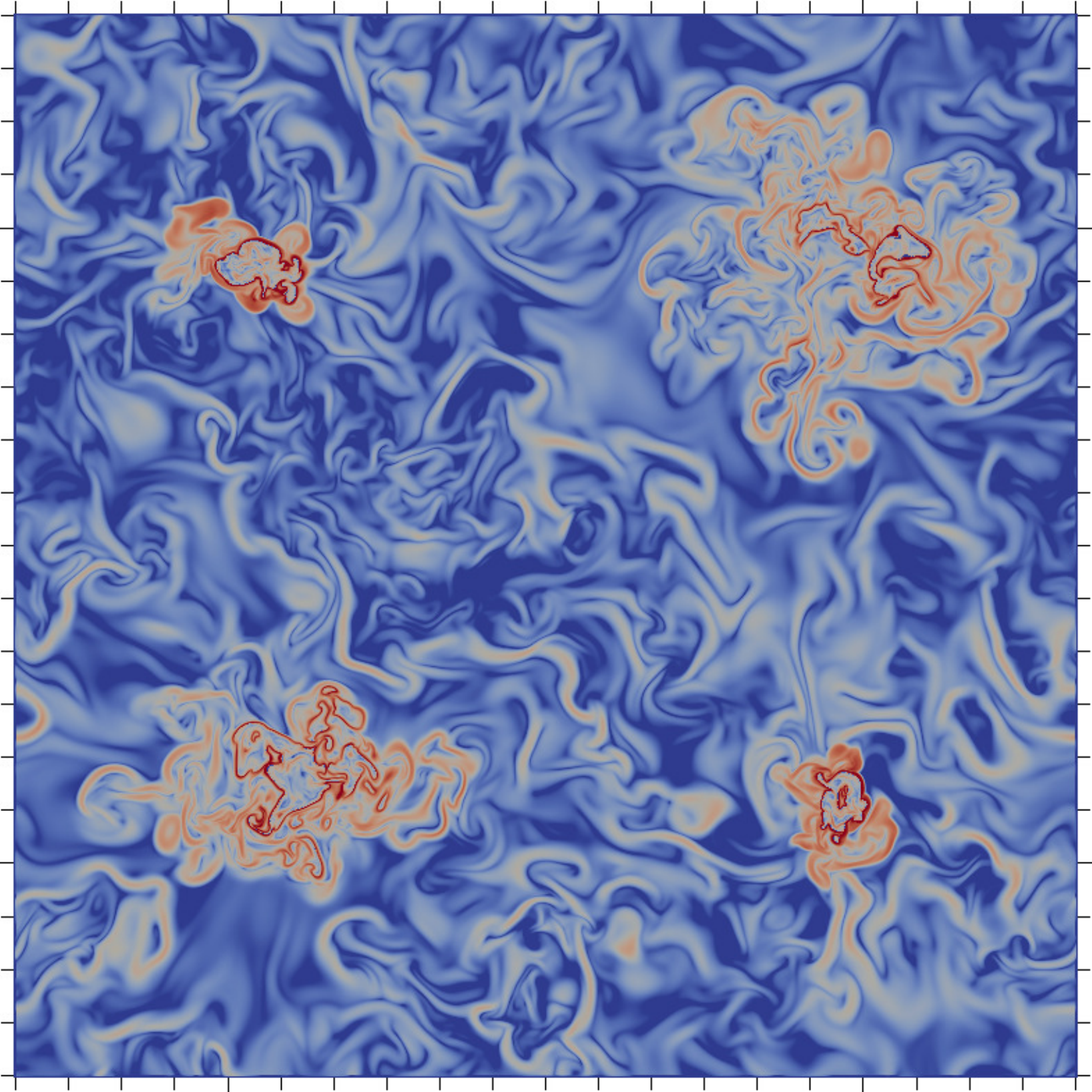}}
    \put( 10, 0){\includegraphics[scale=0.23]{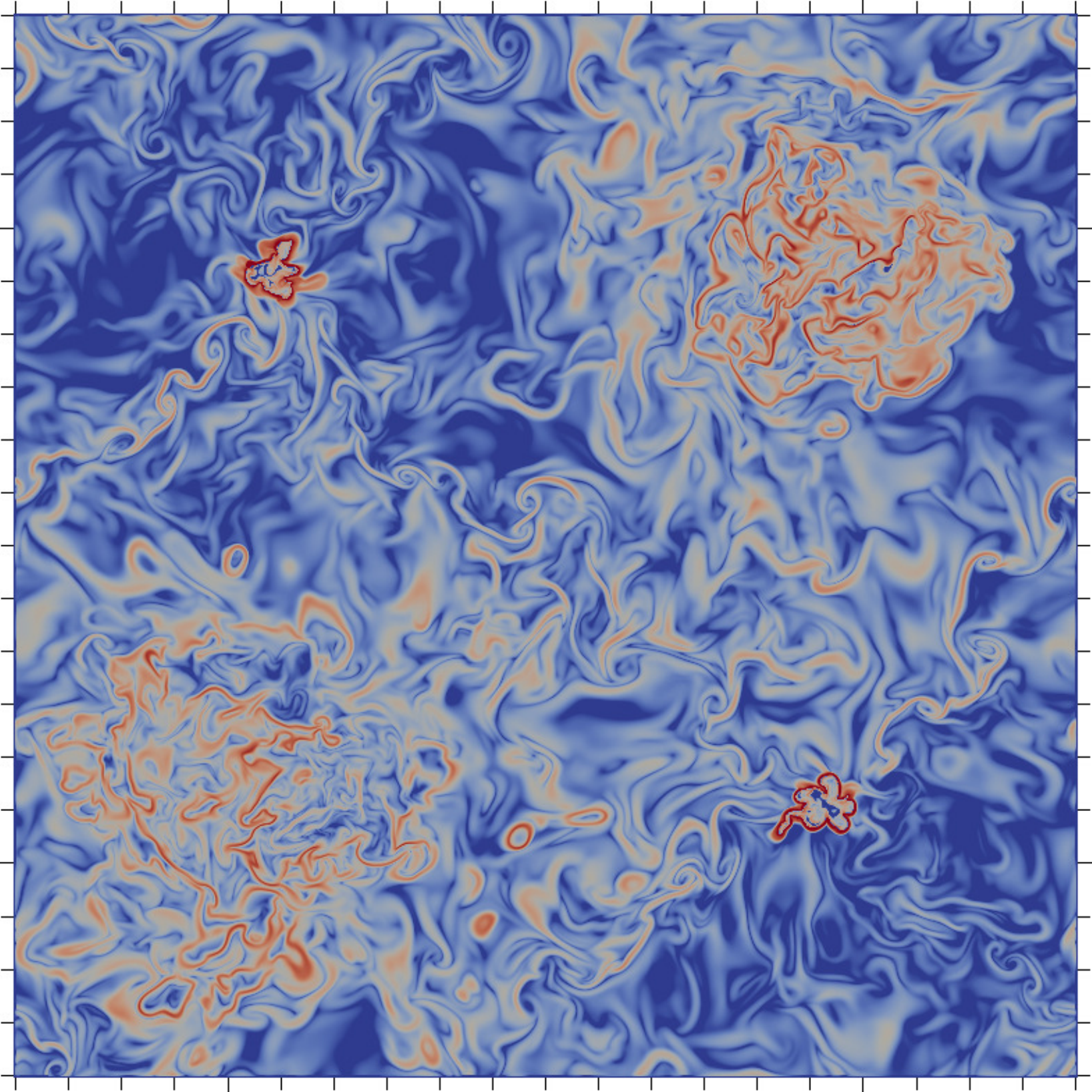}}
    \put( -2,  0){{\footnotesize $0.0$}}
    \put( -2,  44){{\footnotesize$0.5$}}
    \put( -2,  88){{\footnotesize$1.0$}}
    \put( -2,  132){{\footnotesize$1.5$}}
    \put( -2,  176){{\footnotesize$2.0$}}
    \put( -10,  90){$y$}
    \put( 5,  -7){{\footnotesize $0.0$}}
    \put( 50, -7){{\footnotesize$0.5$}}
    \put( 95, -7){{\footnotesize$1.0$}}
    \put( 140, -7){{\footnotesize$1.5$}}
    \put( 185, -7){{\footnotesize$2.0$}}
    \put( 100, -15){$x$}
  \end{picture}
  \end{center}
  \caption{The BPE production
    $\Phi_{d}=(\rd z_{*}/\rd b)|\nabla\,b|^{2}/Pe$ over horizontal
    slices of the domain using a logarithmic scale. The slices
    were taken at $z=0, 0.1, 0.2, 0.45$ clockwise from top
    left. The plumes in each slice can be seen in the top-left and
    bottom-right quadrants of each window and the jets in the
    bottom-left and top-right quadrants. For the colour scale used
    in the figure see figure \ref{fig:vslices}.}
  \label{fig:hslices}
\end{figure}

The BPE production over vertical and horizontal slices is
displayed in figures \ref{fig:vslices} and \ref{fig:hslices},
respectively. These figures illustrate that regions of high
$\Phi_{d}$ typically occur on thin surfaces at the instantaneous
edge of the plumes (cf. figure \ref{fig:buoy}, corresponding to
the same point in time), where both $|\nabla b|$ and the
probability density $\partial z_{*}/\partial b$ are relatively
large because the buoyancy is close to a background buoyancy
$\pm b_{m}$. Diapycnal mixing is energetically insignificant in
the vicinity of the interface outside the plumes for the aspect
ratio $S=1$ shown in figure \ref{fig:vslices}, although the
instantaneous picture of figure \ref{fig:vslices} does not
necessarily account for intermittent events that would increase
the mean BPE production.

The local kinetic energy density $E_{k}$ (per unit mass) is
defined according to $E_{k}\equiv |\vc{u}|^{2}/2$, and satisfies

\begin{equation}
  \pd{E_{k}}{t}+\nabla\cdot (\vc{u}E_{k})=-\nabla\cdot(\vc{u}(p-p_{*}))+\underbrace{w(b-b_{*})}_{-\Phi_{z}} + \frac{1}{Re}\nabla^{2}E_{k}-\varepsilon.
\label{eq:ke}
\end{equation}

\noindent We will refer to the term
$\varepsilon\equiv (\partial_{j}u_{i})^{2}/Re$ as the viscous
dissipation, noting that, prior to spatial integration, the time
average of $\varepsilon$ is strictly only equivalent to the true
viscous dissipation at a given point in homogeneous
turbulence. The physical role played by the relative buoyancy flux
$-\Phi_{z}$ in \eqref{eq:ape} and \eqref{eq:ke} is the reversible
conversion of APE into kinetic energy \citep{WinKjfm1995a}.

\begin{figure}
  \begin{center}
  \begin{picture}(400,200)(0,-15)
    \put( 20,  0){\includegraphics[scale=0.23]{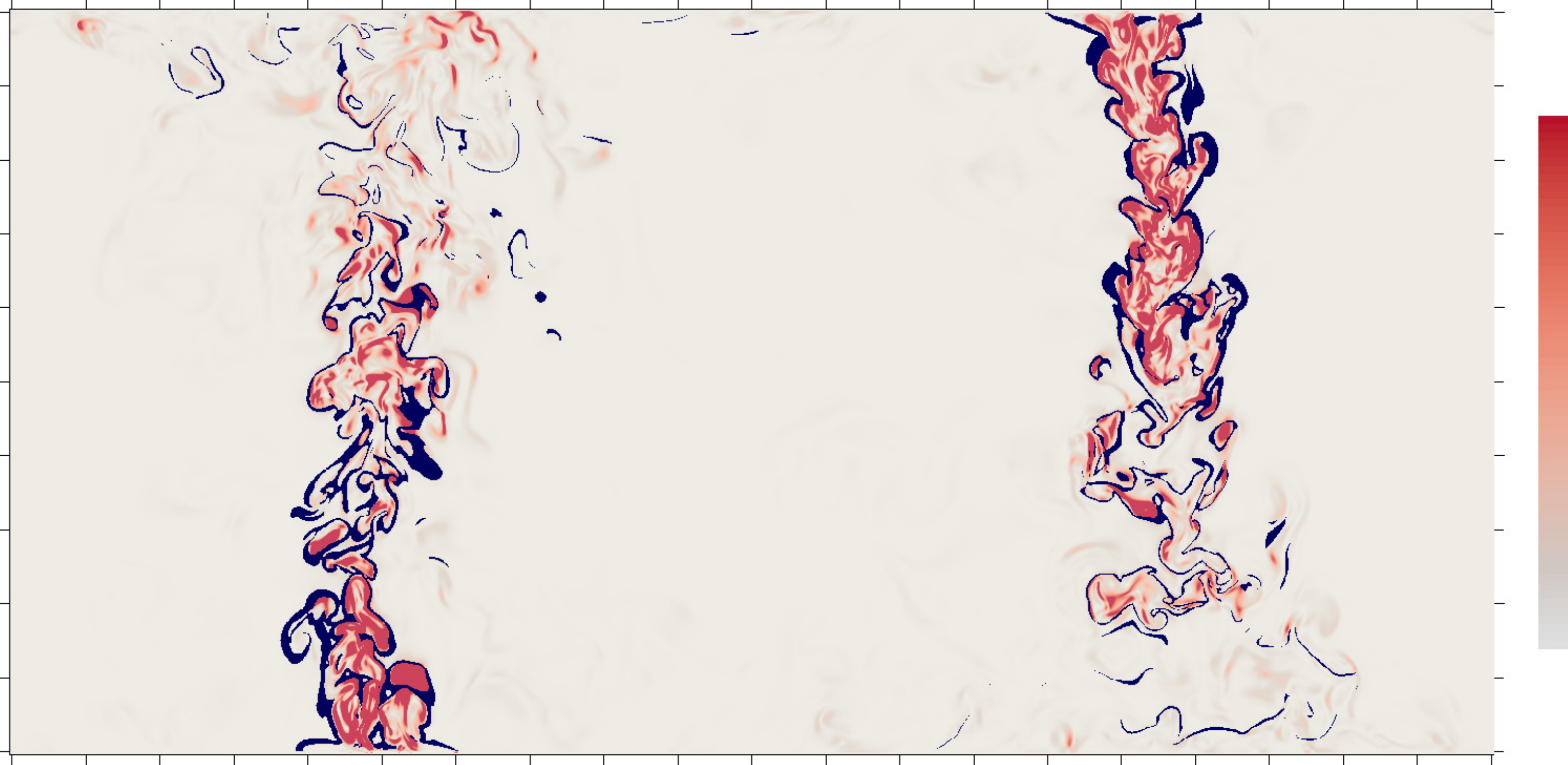}}
    \put( 0,  0){{\footnotesize $-0.5$}}
    \put( 5,  83){{\footnotesize$0.0$}}
    \put( 5,  165){{\footnotesize$0.5$}}
    \put( -5,  90){$z$}
    \put( 20,  -8){{\footnotesize $0.0$}}
    \put( 101,  -8){{\footnotesize $0.5$}}
    \put( 182,  -8){{\footnotesize $1.0$}}
    \put( 264,  -8){{\footnotesize $1.5$}}
    \put( 346,  -8){{\footnotesize $2.0$}}
    \put( 360,  18){$0.001$}
    \put( 360, 146){$10$}
    \put( 192,  -15){$x$}
  \end{picture}
  \end{center}
  \caption{Regions of viscous dissipation $\varepsilon=(\partial_{j}u_{i})^{2}/Re$
    coloured using a logarithmic scale. The thin dark (blue) parts
    of the figure denotes regions in which the BPE production
    $\Phi_{d}=\rd z_{*}/\rd b|\nabla\,b|^{2}/Pe\geq 3$.}
  \label{fig:dis_mix}
\end{figure}

The viscous dissipation $\varepsilon$ of kinetic energy in
\eqref{eq:ke} depends exclusively on local quantities, in the
sense that it is evaluated from properties of the velocity field
at a single point. In contrast, the BPE production $\Phi_{d}$ (and
APE dissipation) depends explicitly on the global probability
density function for buoyancy. As illustrated in figure
\ref{fig:dis_mix}, $\varepsilon$ is relatively large in the core
of the plume, which is enveloped by surfaces on which $\Phi_{d}$
is maximised. Such surfaces correspond to relatively large values
of both $|\nabla b|$ \emph{and} the probability density function
for buoyancy (i.e. surfaces on which $b\approx \pm b_{m}$,
corresponding approximately to the the white lines in figure
\ref{fig:buoy}).

\subsection{Integrals and statistics for energetics}
\label{sec:energy_stats}

The quadrant volume integrals of $\Phi_{z}$, $\varepsilon$ and
$\Phi_{d}$, multiplied by a factor of $2$, are displayed in table
\ref{tab:01}. The factor of $2$ was introduced to ensure that the
theoretical predictions of $\Phi_{z}$, $\varepsilon$ and
$\Phi_{d}$ each correspond to unity. The values indicate that,
whilst substantially more BPE production takes place in the lower
layer than in the upper layer, the distribution of viscous
dissipation in each of the quadrant's layers is approximately
uniform.

Integration of \eqref{eq:ke} over the volume of a
quadrant and time, indicates that the integral of viscous
dissipation is equal to the work done by buoyancy,

\begin{equation}
  \vav{\varepsilon}=-\vav{\Phi_{z}},
\end{equation}

\noindent where $\vav{\Phi_{z}}=-\vav{w(b-b_{*})}=-\vav{wb}$,
because $b_{*}$ does not depend on $x$ or $y$ and therefore
$\vav{wb_{*}}$ is identically zero.   It is interesting that for
the two largest aspect ratios $\vav{\Phi_{z}}$ is greater than
$1$, which exceeds the buoyancy flux provided by the sources. The
cause of the disparity is seen when the governing equation for
buoyancy \eqref{eq:lbuoy} is multiplied by $z$ to give a
volumetric budget for the potential energy $E_{a}+E_{b}=-bz$,
which implies that

\begin{equation}
  -\vav{\Phi_{z}}-\vav{\Phi_{i}}=\underbrace{-\frac{1}{Pe}\Big\langle{ \pd{b}{z}z  }\Big\rangle_{-\zeta}^{\zeta}}_{=1/2},
  \label{eq:phi_z}
\end{equation}

\noindent over a single quadrant. The right-hand side of
\eqref{eq:phi_z} is equal to $1/2$ because a given quadrant
contains a single source of buoyancy at either $z=-1/2$ or
$z=1/2$. The input of potential energy at the boundaries is
therefore equal to the sum of the integral of the convective
buoyancy flux and diffusion of buoyancy down its mean profile
$-\vav{\Phi_{i}}\equiv-\vav{Pe^{-1}\partial_{z}b}$. Indeed, there
is a flux of buoyancy due to the plume in addition to a (negative)
diffusive flux down the stable background
stratification. Together, these contributions result in
$\vav{\Phi_{z}}$ being either slightly higher or slightly lower
than the input at the boundaries, depending on the area of the
plume sources, relative to the area of the horizontal
boundaries. Nevertheless, the contribution from $\vav{\Phi_{i}}$
is small relative to the buoyancy flux $\vav{\Phi_{z}}$, due to
the relatively high P\'{e}clet number of the flow, which results
in a close agreement between $-\vav{\Phi_{z}}$ and the surface
buoyancy fluxes in \eqref{eq:phi_z}.

Unlike the viscous dissipation, which relies on the conversion of
APE to kinetic energy via $\vav{\Phi_{z}}$, the volumetric
production of BPE can be calculated directly from surface fluxes,

\begin{equation}
  \vav{\Phi_{d}}=\underbrace{\frac{1}{Pe}\Big\langle{ \pd{b}{z}z_{*} }\Big\rangle_{-\zeta}^{\zeta}}_{=1/2},
\label{eq:phi_d_unity}
\end{equation}

\noindent for sufficiently long time averages, as is verified to
within approximately $1\%$ by the values (scaled by a factor of
$2$) in table \ref{tab:01}. The data in table \ref{tab:01} also
show that the total input of APE at the boundaries, which is equal
to twice the input of potential energy, is split equally between
BPE production and viscous dissipation, leading to the mixing
efficiency of approximately $1/2$ that is also found in
Rayleigh-B\'{e}nard convection \citep{HugGjfm2013a}.

Table \ref{tab:01} indicates that the viscous dissipation in the
lower layer $\vav{\varepsilon}_{-\zeta}^{0}$ generally decreases
with increasing aspect ratio. In general, and particularly in
domains of small aspect ratio, turbulence is transported
horizontally out of a quadrant's upper layer (which, for the
purpose of discussion, we assume contains the jet) and into
neighbouring quadrants. At larger aspect ratios a greater
proportion of this turbulence is dissipated before it is
transported out of the upper layer, leading to the observed
decrease in lower-layer viscous dissipation for large aspect
ratios.

\begin{table}
  \begin{center}
\def~{\hphantom{0}}
  \begin{tabular}{lcccccccccccc}
    \vspace{2mm}
    $S$ & $-\llangle \Phi_{z}  \rrangle$ & $ -\llangle \Phi_{z}  \rrangle_{-\zeta}^{0}$ & $-\llangle \Phi_{z} \rrangle_{0}^{\zeta}$ & & $\llangle \varepsilon  \rrangle$ & $ \llangle \varepsilon  \rrangle_{-\zeta}^{0}$ & $\llangle \varepsilon  \rrangle_{0}^{\zeta}$ & & $\llangle \Phi_{d}  \rrangle$ & $\llangle \Phi_{d}  \rrangle_{-\zeta}^{0}$ & $\llangle \Phi_{d}  \rrangle_{0}^{\zeta}$ \\
$0.50$ &  0.991 &  0.827 &  0.164 &  &  0.979 &  0.584 &  0.395 &  &  1.008 &  0.781 &  0.227 \\
$0.58$ &  1.006 &  0.839 &  0.167 &  &  1.020 &  0.593 &  0.428 &  &  1.005 &  0.780 &  0.225 \\
$0.62$ &  0.982 &  0.870 &  0.111 &  &  0.981 &  0.574 &  0.407 &  &  1.014 &  0.801 &  0.213 \\
$0.67$ &  0.992 &  0.899 &  0.094 &  &  0.990 &  0.535 &  0.455 &  &  1.003 &  0.758 &  0.244 \\
$1.00$ &  1.003 &  0.918 &  0.085 &  &  1.005 &  0.492 &  0.512 &  &  0.997 &  0.738 &  0.259 \\
$1.33$ &  1.005 &  0.923 &  0.082 &  &  1.004 &  0.459 &  0.545 &  &  0.998 &  0.751 &  0.247 \\\\ 
Theory &  1.000 &  1.000 &  0.000 &  &  1.000 &  0.500 &  0.500 &  &  1.000 &  0.750 &  0.250 \\ 

  \end{tabular}
  \caption{Domain decomposition of the buoyancy flux
    $\vav{\Phi_{z}}$, viscous dissipation $\vav{\varepsilon}$ and
    BPE production $\vav{\Phi_{d}}$.
    $\vav{}, \vav{}_{-\zeta}^{0}$ and $\vav{}_{0}^{\zeta}$ refer
    to volume integrals over an entire quadrant, the (lower) plume
    layer and the (upper) jet layer, respectively. The row
    labelled `Theory' corresponds to the results obtained for
    Gaussian profiles in \S\ref{sec:ke_model} and
    \S\ref{sec:ape_model}. Noting that each layer of the domain
    has height $0.5$, the volume integrals displayed in this table
    have been multiplied by a factor of $2$, to facilitate a
    comparison with the horizontal axis of figure \ref{fig:05} and
    the interpretation of their ratios.}

  \label{tab:01}
  \end{center}
\end{table}

Figure \ref{fig:05} displays horizontal integrals of $(a)$ the BPE
production $\hav{\Phi_{d}}$ and $(b)$ the viscous dissipation
$\hav{\varepsilon}$ over the quadrant as a function of
$z$. Consistent with the information in table \ref{tab:01}, figure
\ref{fig:05} indicates that the BPE production is significantly
higher in the plume below the interface than it is in the jet
above the interface, in which it decays with respect to $z$. In
the lower layer the BPE production exhibits a local maximum of
approximately $0.8$ for the largest aspect ratio. Due to the
finite area of the source, the production of BPE production
increases significantly in the vicinity of the bottom boundary. In
the following sections we develop the integral model whose
predictions are compared to the observations in figure
\ref{fig:05}.

\subsection{Viscous dissipation}
\label{sec:ke_model}

In this section we will develop a bulk model for the volume
integral of viscous dissipation $\vav{\varepsilon}$ in the
quadrant layer that contains a jet and the quadrant layer that
contains a plume. We will also describe a model for the vertical
variation of horizontally integrated viscous dissipation in each
layer.

For generality we will assume that the distance from the plume
source to the interface is $0\leq\zeta\leq 1$, and not necessarily
equal to $1/2$.  In practice, the situation we have in mind might
correspond to an `emptying filling box' model of a space
ventilated by low and high-level openings \citep{LinPjfm1990a} or
to a domain containing an unequal number of buoyancy sources on
its bottom boundary compared with its top boundary.  For
consistency with the previous sections, we continue to regard the
ambient buoyancy of the upper layer as $b_{m}$ and that of the
lower layer as $-b_{m}$.

In the absence of a direct input of kinetic energy at the
boundaries, the viscous dissipation in the domain is balanced by
the total work undertaken by buoyancy. Buoyancy is only able to do
work on the flow in the quadrant's lower layer ($z < 0$), because
in the upper layer the mean buoyancy difference between the plume
and the ambient is zero.  As discussed in
\S\ref{sec:energy_stats}, provided that the vertical buoyancy
transport by diffusion is small compared with the relative
buoyancy flux $w(b-b_{*})$, the kinetic energy dissipation is
equal to the buoyancy flux multiplied by $\zeta$, which is the
dimensionless distance over which buoyancy is able to do work in a
single quadrant.

\begin{figure}
  \begin{center}
    \includegraphics[scale=1]{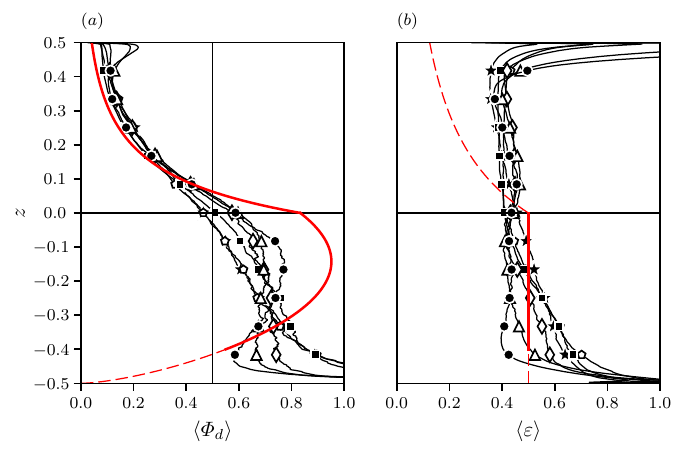}
  \end{center}
  \caption{$(a)$ BPE production and $(b)$ viscous dissipation over
    a quadrant of the domain. In this quadrant the plume occupies
    the lower layer and the jet occupies the upper layer. The
    thick (red) lines correspond to the theoretical
    predictions. The model in $(a)$ is developed in
    \S\ref{sec:similarity}. The model is dashed in regions close
    to the source, because it was developed under the assumption
    of point sources of buoyancy flux. The thick line in $(b)$ is
    the total dissipation in the lower layer
    \eqref{eq:epsilon_lower}, divided by the layer height, which
    gives $\hav{\varepsilon} = 1/2$ for a Gaussian velocity
    profile.  The dashed line in $(b)$ is described by equation
    \eqref{eq:epsjet} and corresponds to viscous dissipation in an
    unconfined jet. For the symbols used to denote each aspect
    ratio see table \ref{tab:02}.}
  \label{fig:05}
\end{figure}

Assuming that the aspect ratio is sufficiently large and that the
upper layer is not stratified, all the kinetic energy that is 
transported across the interface is dissipated in the upper layer. 
In other words, the total kinetic energy dissipation in the layer 
containing the jet is equal to the transport of kinetic energy 
across the interface. 

Using the steady-state plume solutions \specialeqref{eq:QM}{}, and noting
that all quantities are non-dimensionalised using the source
buoyancy flux and domain height, the transport of kinetic energy across
the interface at $z=0$ is

\begin{equation}
 \left.\Big\langle\frac{w^{3}}{2}\Big\rangle\right|_{z = 0} = \frac{\gamma\,M_{m}^{2}}{2Q_{m}}=\frac{3\gamma}{8}\zeta,
\label{eq:ke_flux}
\end{equation}

\noindent where $\gamma$ is therefore a parameter that accounts
for the shape of the velocity profile and turbulent transport
\citep[see, e.g.][]{PriCqms1955a, ReeMjfm2015a}. Note that
equation \eqref{eq:ke_flux} relates to the kinetic energy budget
for which the forcing from the sources of buoyancy is known, and
therefore does not involve the entrainment coefficient
$\alpha$. For the same reason equation \eqref{eq:ke_flux}, unlike
the volume flux $Q$, would not be directly affected by the
division of a plume into multiple plumes that provide the same
combined buoyancy flux.

The total kinetic energy dissipation in the layer containing the plume 
is equal to the integral of the buoyancy flux over the layer, 
minus the transport of kinetic energy between the top and 
bottom layers; hence

\begin{equation}
\vav{\varepsilon}_{-\zeta}^{0}\equiv\int_{-\zeta}^{0}\hav{\varepsilon}\rd z=\zeta\left(1-\frac{3\gamma}{8}\right).
\label{eq:epsilon_lower}
\end{equation}

\noindent This result can also be obtained from the plume
equations by integrating the production term in the integral
energy transport equation \citep[see, e.g.][]{ReeMjfm2015a}.
Equations \eqref{eq:ke_flux} and \eqref{eq:epsilon_lower} state 
that when $\gamma=4/3$, which corresponds to a Gaussian velocity 
profile \citep{ReeMjfm2015a}, there is equal viscous dissipation 
in the upper layer compared with the lower layer.

In the plume, self-similarity implies that the approximately
constant integral buoyancy force doing work on the flow results in
the horizontal integral of viscous dissipation being constant. The
viscous dissipation in the plume for $z\leq 0$ is therefore equal
to $1-3\gamma/8=1/2$ for a Gaussian velocity profile, as indicated
by the red solid line in the lower half of figure
\ref{fig:05}$(b)$. In contrast, fluid in a jet is not subjected to
a mean force from buoyancy, which results in its viscous
dissipation decaying with respect to $z$.  Equating the jet's
dissipation with the vertical derivative of its energy flux
$\gamma\,M^{2}/2Q$ using \eqref{eq:Qjet} implies that

\begin{equation}
  \hav{\varepsilon}=\frac{3\gamma}{8}\frac{\zeta^{2}}{(\zeta+z)^{2}},\ \ z\geq 0.
  \label{eq:epsjet}
\end{equation}

We include \eqref{eq:epsjet} as a dashed curve in the upper layer
of figure \ref{fig:05}$(b)$ as a reference to the behaviour of an
unconfined jet. We note, however, that \eqref{eq:epsjet} does not
give accurate predictions in the vicinity of the top boundary due
to vertical confinement. Indeed, the value
$2\,\vav{\varepsilon}_{0}^{\zeta}=0.5$ in table \ref{tab:01}
can only be obtained by integrating \eqref{eq:epsjet} from $z=0$
to $\infty$.

\subsection{An integral model for the production of BPE}
\label{sec:ape_model}

In this section we will develop a model for the integral
production of BPE in the layer containing the jet and the layer
containing the plume. As discussed in \S\ref{sec:energy}, our
model does not distinguish between BPE production $\Phi_{d}$ and
APE dissipation $G$, which readers can regard as interchangeable in
the following model, in spite of the significant differences
between $\Phi_{d}$ and $G$ in more general cases \citep[see, for
example,][]{TaiRjfm2009a}. In a stable two-layer stratification
consisting of an upper and lower layer of buoyancy $b_{m}$ and
$-b_{m}$, respectively, the sorted elevation of fluid parcels is

\begin{equation}
  z_{*}(b)=
\begin{cases}
  1-\zeta,\quad\quad b_{m}<b, \\
  0,\quad\quad -b_{m}<b<b_{m}, \\
  -\zeta,\quad\quad b<-b_{m}.
\end{cases}
  \label{eq:zs}
\end{equation}

\noindent Parcels whose buoyancy is exactly equal to $-b_{m}$ and
$b_{m}$ occupy the regions $-\zeta<z_{*}<0$ and $0<z_{*}<1-\zeta$,
respectively. Using \eqref{eq:Ea2}, the local APE can be evaluated as

\begin{equation}
  E_{a}(b,z) = b(z_{*}-z)+b_{m}(|z|-|z_{*}|),
  \label{eq:Ea_layered}
\end{equation}

\begin{figure}
  \begin{center}
    \includegraphics[scale=1]{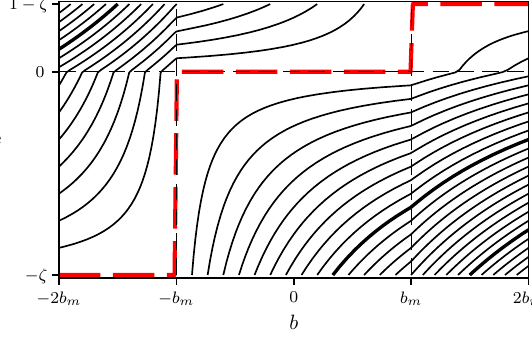}
  \end{center}
  \caption{Contours of the local available potential energy $E_{a}$ according to
    \eqref{eq:Ea_layered}, with respect to the buoyancy and
    vertical position for parcels displaced from a stable
    two-layer stratification. The thick dashed line corresponds to
    $z_{*}$, which minimises $E_{a}$. The spacing of the contours
    is equal to $0.1$.}
\label{fig:psi}
\end{figure}
\begin{figure}
  \begin{center}
    \includegraphics[scale=1]{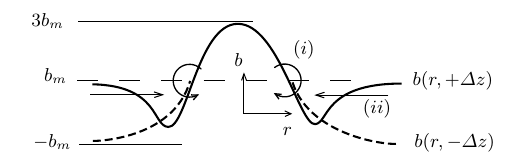}
  \end{center}
  \caption{Schematic description of the mean buoyancy profile a
    small distance above ($\Delta z$) and below ($-\Delta z$) the
    interface. When the plume crosses the interface between the
    two layers it moves from a layer with mean buoyancy $-b_m$ to
    a layer with mean buoyancy $b_m$. The buoyancy profile changes
    from $b(r,-\Delta z)$ (thick dashed line) to $b(r,+\Delta z)$
    (thick solid line). The core of the jet is positively buoyant
    relative to the layer buoyancy, shown by the central part of
    the solid line being above $b_m$, while the outer shell of the
    plume is negatively buoyant relative to the layer buoyancy. As
    the jet travels through the layer, the radial buoyancy
    variation in the jet is mixed out. We model this as $(i)$
    mixing between the core and the outer shell and $(ii)$
    entrainment from outside the jet into the outer shell. In
    \S\ref{sec:ape_transport} we quantify the role of turbulent
    fluctuations that modify this simple heuristic picture.}
  \label{diag:08}
\end{figure}


\noindent whose contours are displayed in figure
\ref{fig:psi}. Note that $E_{a}(b_{m},z)=0$ for $z>0$ and
$E_{a}(-b_{m},z)=0$ for $z<0$. Conceptually \eqref{eq:Ea_layered}
states that APE comes either from
negatively buoyant parcels of fluid displaced upwards from $z<0$
or from positively buoyant parcels of fluid displaced downwards
from $z>0$, as illustrated either side of the interface in figure
\ref{diag:08}. For arbitrary interface heights $\zeta$, $E_{a}$ is
not symmetric around the interface, because the vertical distance
to equilibrium for parcels of fluid with $b<-b_{m}$ is not
necessarily the same as it is for parcels with $b>b_{m}$.

Multiplication of equation \eqref{eq:Ea_layered} by $w$ and
integration over a horizontal slice (i.e. application of
$\hav{\cdot}$) gives an expression for the flux of APE in the
jet. Noting that $\hav{\Phi_{z}}=-\hav{w(b-b_{*})}=0$ when $z>0$
(because in the upper layer the jet has the same mean buoyancy as
the ambient), the flux can be expressed as

\begin{equation}
  \hav{w E_{a}}=\frac{1-\zeta}{2}\hav{|\Phi_{z}|},
\label{eq:wEa}
\end{equation}

\noindent where

\begin{equation}
\frac{\hav{|\Phi_{z}|}}{2}=\int_{b>b_{m}}w(b-b_{m})\mathrm{d}x\mathrm{dy},
\end{equation}

\noindent corresponds to the quadrant integral of the flux
$\Phi_{z}$ over regions where $b<b_{m}$.

We proceed by assuming that for domains of sufficiently large
aspect ratio, the ambient is well-mixed and of uniform buoyancy
and, therefore, that the horizontal flux of APE through the sides
of the quadrant is equal to zero. Our observations of the mean
horizontal flux of APE out of a quadrant indicate that it is
negligible for domains with aspect ratio $S\geq 1$.  If there is
no transport of APE out of a quadrant, the APE flux
$\hav{w E_{a}}$ at the the level of the interface is equal to the
total dissipation of APE in the upper layer. Note that in the
upper layer the buoyancy flux $-\hav{\Phi_{z}}$ is zero; hence the
overall conversion of APE to kinetic energy is also zero:

\begin{equation}
  \vav{G}_{0}^{1-\zeta}\approx\vav{\Phi_{d}}_{0}^{1-\zeta}=\int_{0}^{1-\zeta}\hav{\Phi_{d}}\rd z = (1-\zeta)\beta,
\label{eq:phi_d}
\end{equation}

\noindent where

\begin{equation}
  \beta \equiv \frac{\hav{|\Phi_{z}|}}{2}\bigg|_{z=0}
  \label{eq:beta_definition}
\end{equation}

\noindent is a parameter that depends on the shape of the buoyancy
and velocity profiles at the interface and, relating to a buoyancy
flux, does not depend explicitly on entrainment. As discussed in
\S\ref{sec:energy} the reasoning leading to \eqref{eq:phi_d}
assumes that the P\'{e}clet number is sufficiently large for this
Boussinesq flow to equate BPE production $\hav{\Phi_{d}}$ with APE
dissipation $G$.

Buoyancy can only do work on the flow over a height $\zeta$;
therefore, not more than $\zeta$ of the supply of APE from the
plume source is converted into kinetic energy, which means that
the total diapycnal mixing in the lower layer is

\begin{equation}
  \vav{\Phi_{d}}_{-\zeta}^{0}=1 - \vav{\Phi_{d}}_{0}^{1-\zeta} - \zeta=(1-\zeta)(1-\beta).
\label{eq:phid_lower}
\end{equation}

\noindent If the profiles for velocity and buoyancy in the plume
are assumed to have a Gaussian form and temporal fluctuations are
neglected then $\beta = 1/4$, as demonstrated by equation \eqref{eq:Phiz_above} in appendix
\ref{sec:beta}. Consequently, for $\zeta=1/2$, equation
\eqref{eq:phid_lower} predicts that $3/8$ of the APE is converted
into BPE in the lower layer, while $1/8$
of the APE is converted into BPE in the
upper layer. As can be seen in the final two columns of table
\ref{tab:01}, this prediction is consistent with observations from
the domain with $S=4/3$ to within less than $1\%$. The remaining
$1/2$ of the supply of APE is converted into kinetic energy, as
outlined in the previous section.

As supported by the observations reported in table \ref{tab:01},
the mixing efficiency of this system, defined as
$\vav{\Phi_{d}}/(\vav{\varepsilon}+\vav{\Phi_{d}})$
\citep{HugGjfm2013a} is equal to $1/2$ when $\zeta=1/2$. For other
values of $\zeta$, the mixing efficiency would be equal to
$1-\zeta$, which is consistent with the analysis of an emptying
filling box \citep{WykMarx2018a}. We note that $\zeta\neq 1/2$
could be manufactured in a quadrant of a closed domain by dividing
the buoyancy flux from a single point source between multiple
point sources \citep{LinPjfm1990a} in one of the layers. However,
if the mixing efficiency were then calculated over adjacent
quadrants it would always be $1/2$, which suggests that the
efficiencies that differ from $1/2$ in \citet{WykMarx2018a} are
associated with the open nature of the emptying filling box
domain.

\subsection{A model for the vertical dependence of BPE production}
\label{sec:similarity}

In this section we will derive a model for the scaling 
of the vertical flux of APE $\hav{wE_{a}}$ with height and, 
consequently, the BPE production as a function of height. 
As outlined in the previous sections, the model is
based on two layers of ambient fluid of infinite horizontal
extent, leading to the reduced form of \eqref{eq:ape_budget},

\begin{equation}
\od{}{z}\hav{wE_{a}}=-\hav{\Phi_{d}}+\hav{\Phi_{z}},
\label{eq:ape_1d}
\end{equation}

\noindent which states that the destruction of APE
$\hav{G}\approx\hav{\Phi_{d}}$ balances the work done by buoyancy
$\hav{\Phi_{z}}$ and differences in the boundary fluxes of
APE.

Using Gaussian profiles for velocity and buoyancy, and
\eqref{eq:Ea_layered} to evaluate $E_{a}$ leads to the following
expression for the APE flux in the lower
layer (see appendix \ref{sec:beta}):

\begin{equation}
  \hav{wE_{a}}=1-\zeta\,Z-(1-\zeta)Z^{5/3}+\frac{1-\zeta}{4}Z^{10/3},\ \ 0\leq Z < 1,
\label{eq:wEa_below_main}
\end{equation}

\noindent where $Z\equiv (z + \zeta)/\zeta$ is a rescaled vertical
coordinate that measures the distance from the source as a
proportion of the distance from the plume source to the interface
($Z = 0$ at the source and $Z = 1$ at the interface). In the lower
layer, the relative buoyancy flux $-\hav{\Phi_{z}}$ is unity;
hence, noting that $\zeta\rd Z=\rd z$, \eqref{eq:ape_1d} indicates
that

\begin{equation}
\hav{\Phi_{d}}=\frac{5}{6}\frac{1-\zeta}{\zeta}\left(2Z^{2/3}-Z^{7/3}\right),\ \ 0\leq Z < 1.
\label{eq:Phi_d_lower}
\end{equation}

\noindent At the interface, where $Z=1$, \eqref{eq:Phi_d_lower}
implies that

\begin{equation}
\hav{\Phi_{d}}\bigg|_{Z=1}=\frac{5}{6}\frac{1-\zeta}{\zeta}.
\end{equation}

In the upper layer, the APE dissipation and BPE production are
equal to (minus) the vertical derivative of the APE flux:

\begin{equation}
  \hav{G}=\hav{\Phi_{d}}=-\frac{1-\zeta}{2}\od{\hav{|\Phi_{z}|}}{z}>0,\quad z>0,
\label{eq:phi_d2}
\end{equation}

\noindent using \eqref{eq:wEa}. Progress beyond \eqref{eq:phi_d2}
requires a model for the evolution of the APE flux in the upper
layer, which depends on the rate at which the positively and
negatively buoyant regions in the jet mix, as illustrated
schematically in figure \ref{diag:08}. We develop a heuristic
description of the process in terms of the mean buoyancy by
assuming that the mixing of the core and the annular shell of the
jet $(i)$ proceeds a rate that is proportional to the buoyancy
deficit $b-b_{m}$. At the same time, the buoyancy deficit is
reduced by $(ii)$ the overall dilution of the jet due to entrained
ambient fluid. The dilution is accounted for by noting that
$b-b_{m}$ is proportional to the flux of APE,
$(1-\zeta)\hav{|\Phi_{z}|}/2$, divided by a volume flux.

Provided that the virtual source of the jet coincides with that of
the plume at $z=-\zeta$, the volume flux in the jet is
proportional to $z+\zeta$, which is true for Gaussian jets and
plumes because their spreading rates are almost identical
\citep[see equation \eqref{eq:Qjet} and][]{ReeMjfm2015a}; hence

\begin{equation}
\od{\hav{|\Phi_{z}|}}{z}=-c\frac{\hav{|\Phi_{z}|}}{z+\zeta},\quad z>0.
\label{eq:phi_d_model}
\end{equation}

The processes of $(i)$ mixing and $(ii)$ overall dilution
described above, and depicted in figure \ref{diag:08}, both relate
to entrainment. Yet, as outlined at the start of this section, we
do not expect an integral energy budget to require a closure for
entrainment. With regards to $(i)$, it is known that the radial
component of the turbulent transport of buoyancy is proportional
to an entrainment coefficient in self-similar jets (in the case of
passive scalars) and in plumes \citep{CraJjfm2016a}. However, the
entrained volume that is responsible for suppressing the rate at
which process $(i)$ occurs is also proportional to an entrainment
coefficient. Processes $(i)$ and $(ii)$ therefore combine in
\eqref{eq:phi_d_model} to produce the parameter $c$ that can be
deduced without explicit reference to an entrainment coefficient.

Using \eqref{eq:phi_d_model}, \eqref{eq:phi_d2} and ensuring
$C^{0}$ continuity of $\hav{\Phi_{d}}$ at $z=0$, implies that

\begin{equation}
  c=\frac{5}{3\zeta}. \label{eq:c1}
\end{equation}

\noindent Integration of \eqref{eq:phi_d_model} and substitution of the result into \eqref{eq:phi_d2} yields

\begin{equation}
\hav{|\Phi_{z}|} = \frac{1}{2Z^{c}},\quad
\mathrm{and}\quad \hav{\Phi_{d}} = \frac{5}{12}\frac{1-\zeta}{\zeta^{2}}\frac{1}{Z^{c+1}}.
\label{eq:model}
\end{equation}

\noindent For an interface at a distance $\zeta = 1/2$ from the source, 
\eqref{eq:model} implies that

\begin{equation}
\hav{|\Phi_{z}|} = \frac{1}{2Z^{10/3}},\quad
\mathrm{and}\quad \hav{\Phi_{d}} = \frac{5}{6Z^{13/3}},\ \ Z\geq 1.
\label{eq:gaussian_model}
\end{equation}

\noindent The predictions obtained from \eqref{eq:Phi_d_lower}
(lower layer, $0\leq Z< 1$) and \eqref{eq:gaussian_model} (upper
layer, $1\leq Z$) are included in figure \ref{fig:05}$(a)$. The
upper layer model exhibits a reasonably good agreement with the
simulation data for large aspect ratios, although the predicted
value of $\Phi_{d}$ at the interface is relatively large. Indeed,
the model assumed a step change in the ambient buoyancy rather
than the more gradually varying profiles of buoyancy that are seen
in the simulations (cf. figure \ref{fig:01}). It should also be
noted that the model does not account for mixing induced by the
horizontal boundaries because it does not include a length scale
corresponding to the vertical confinement.

There appears to be a large difference between the observations
and the model's predictions in the lower layer. However, the
observations of $\Phi_{d}$ depend on aspect ratio and there are
indications that the simulations exhibit an increasingly good
agreement with the model's prediction as the aspect ratio
increases. It is also worth noting that the model assumes a point
source of buoyancy, rather than the finite circular source that
was used in each simulation, leading to large differences between
prediction and observation close to the source. On the largest
domain, the existence of a local maximum in $\hav{\Phi_{d}}$ at
approximately $z=-0.2$ is consistent with the model. We therefore
suspect that observations of plumes emanating from smaller sources
on domains of larger aspect ratio than those studied here would
yield an improved agreement with the model.

\subsection{The radial dependence of velocity and buoyancy}
\label{sec:rprofs}

From an integral perspective, the heuristic arguments leading to
the scaling described in \S\ref{sec:similarity} are reasonably
consistent with the integral observations shown in figure
\ref{fig:05}. However, the extent to which the integral model
faithfully accounts for the underlying transport processes is
unclear and is therefore the subject of this and the subsequent
section, in which we study the radial dependence and vertical
evolution of time and azimuthally averaged quantities.

Azimuthal averages were obtained by identifying computational
cells that lie an equal distance from the vertical centre line of
the two jets or plumes in a given layer (see, for example, the
coordinates labelled `$r$' in figure \ref{diag:01}). We start by
comparing the data to mean Gaussian profiles for velocity and
buoyancy, derived in appendix \ref{sec:beta}, whose predicted amplitude and
width scale according to the model described in
\S\ref{sec:similarity}. We note, however, that such a comparison
is naive in relying on the assumption that the APE flux used in
\S\ref{sec:similarity} is comprised entirely from mean flow
processes, which we will demonstrate is not the case in
\S\ref{sec:ape_transport}.

\begin{figure}
  \begin{center}
    \includegraphics[scale=1]{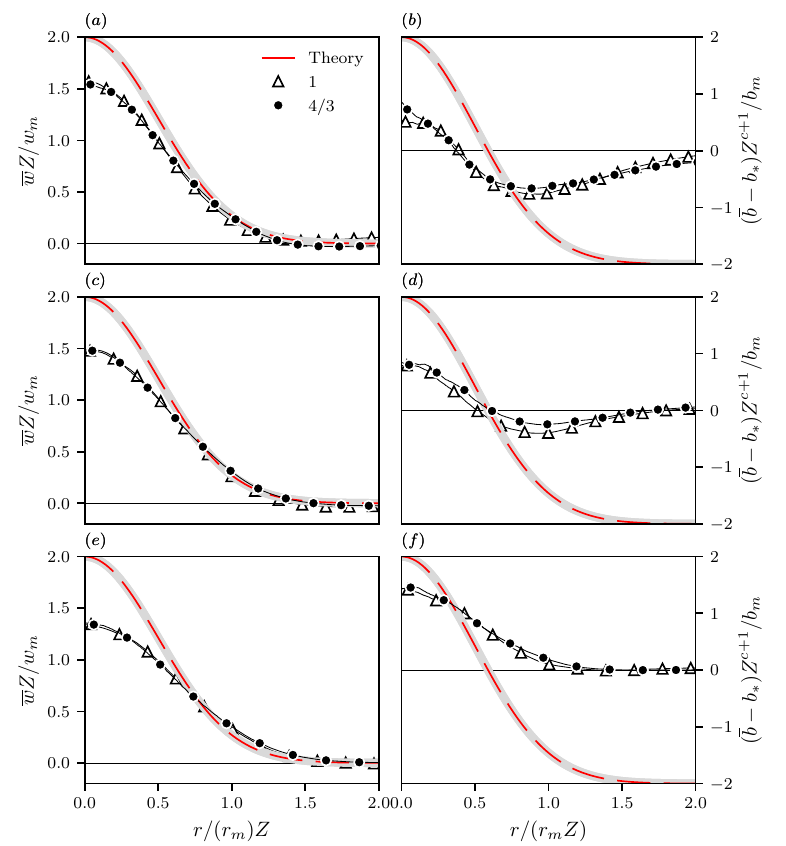}
  \end{center}
  \caption{Mean velocity $\av{w}$ (left column) and mean buoyancy
    in the jet, relative to the background state, $\av{b}-b_{*}$
    (right column) with respect to the radial similarity
    coordinate $r/(r_{m}Z)$, where $r_{m}=6\alpha\zeta/5$ is the
    radius of the plume at the interface and $Z$ is the distance
    from the source rescaled by the distance from the plume source
    to the interface $Z=(z + \zeta)/\zeta$ ($= 2z+1$ for
    $\zeta = 1/2$) at heights $z=0.3\ (a,b)$, $z=0.1\ (c,d)$ and
    $z=0.0\ (e,f)$. The symbols denote data corresponding to
    aspect ratios $S=1$ and $S=4/3$.}
  \label{fig:26}
\end{figure}

Figure \ref{fig:26} displays the radial dependence of the mean
vertical velocity $\av{w}$ $(a,c,e)$ and mean buoyancy $\av{b}$
$(b,d,f)$ in the upper layer at heights $z=0.3\ (a,b)$ $z=0.1\ (c,d)$ and
$z=0.0\ (e,f)$ on domains of aspect ratio $1$ and $4/3$. The
observed profiles in figure \ref{fig:26} are rescaled using the
model's predictions in terms of the rescaled distance 
$Z=(z + \zeta)/\zeta$ from the source, rather than being scaled 
arbitrarily to produce the best
local agreement. The comparison therefore tests the ability of the
model to predict the shape of the profiles, in addition to their
evolution in the vertical direction.

The panels $(a), (c)$ and $(e)$ indicate that the velocity in the
jet after the plume penetrates the stratification has an
approximately Gaussian dependence on the radial coordinate. The
similarity of the profiles in each panel suggest that the spatial
evolution of $\av{w}$ is indeed self-similar, and therefore
confirms the explanation that was provided in \S\ref{sec:flow},
regarding the development of secondary circulation cells due to
entrainment into the jets.

The model consistently over-estimates the maximum velocity, which
is partly due to assignment of the total buoyancy flux input
exclusively to mean transport, rather than turbulent
transport. Whilst it is possible to account for the effects of
turbulent transport in plume theory \citep{CraJjfm2016a}, we
refrain from doing so here to keep our approach as transparent as
possible. It should also be noted that integrals over the
horizontal cross-section of the flow are relatively insensitive to
azimuthally averaged values close to the centre line ($r=0$).

Relative to its local environment, the observed mean buoyancy
displayed in \ref{fig:26}$(b,d,f)$ is positive in the core of the
jet and negative in a shell starting at $r/(r_{m}Z)\gtrapprox 0.5$
for heights $z=0.1$ and $z=0.3$, where $r_{m}=6\alpha\zeta/5$ is
the radius of the plume at the interface. However, the amplitude
of the observed mean buoyancy anomalies is significantly smaller
than that predicted by theory. At the height of the interface
($z=0.0$), the model predicts a step change in the ambient
buoyancy, whereas the observed ambient buoyancy varies gradually
(cf. figure \ref{fig:01}). In this respect it is interesting that
the observed mean buoyancy profile in panel $(f)$ is nowhere less
than the ambient buoyancy at the interface. The buoyancy profile
therefore only evidences the effects of a changing ambient for
larger values of $z$. Unlike the mean velocity profiles, the
average buoyancy profiles do not appear to be self-similar. For
example, the shell of negatively buoyant fluid in $(b)$ at $z=0.3$
is relatively large in comparison with the equivalent region in
panel $(d)$ at $z=0.1$.

It should be noted that the model developed in
\S\ref{sec:similarity} does not permit a unique interpretation of
the way in which the mean buoyancy returns to its ambient value
when $r/(r_{m}Z)\rightarrow \infty$ at a given elevation. Outside
the jet, the vertical velocity is negligible. Therefore, the mean
buoyancy flux, which is the primary quantity on which the model is
based, is also negligible, regardless of the precise form of the
buoyancy profile.

The difference between theory and observation in the mean buoyancy
profile shown in figure \ref{fig:26} raises the question of how
the theory is able to correctly describe the integral scaling of
BPE production in the upper layer of figure \ref{fig:05}$(a)$. The
prediction of BPE production requires an accurate description of
APE transport, whose constituent physical mechanisms and radial
distribution we therefore study in the following section.

\begin{figure}
  \begin{center}
    \includegraphics[scale=1]{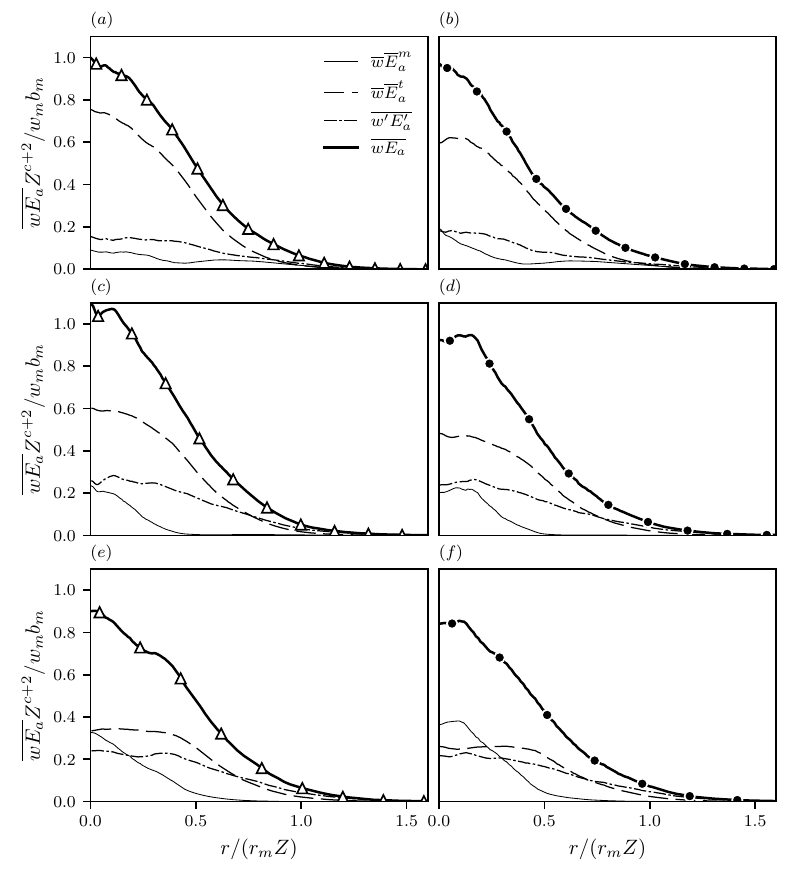}
  \end{center}
  \caption{Observed profiles of the mean advective transport of
    APE $\av{wE_{a}}=\av{w}\av{E}_{a}^{m}+\av{w}\av{E}_{a}^{t}+\av{w'E_{a}'}$
    with respect to the radial similarity coordinate $r/(r_{m}Z)$,
    where $Z$ is the distance from the source rescaled by the
    interface height $Z=(z + \zeta)/\zeta$ ($= 2z+1$ as
    $\zeta = 1/2$) at heights $z=0.3\ (a,b)$, $z=0.1\ (c,d)$ and
    $z=0.0\ (e,f)$. The left column corresponds to the aspect
    ratio $S=1$ and the right column corresponds to the aspect
    ratio $S=4/3$.}
  \label{fig:25}
\end{figure}

\subsection{The turbulent transport of APE}
\label{sec:ape_transport}

Subject to the assumptions laid out in \S\ref{sec:ape_model}, the
prediction of BPE production in a given control volume requires
knowledge of boundary fluxes of APE. In this regard, the model
described in \S\ref{sec:ape_model} was motivated heuristically in
terms of mean profiles of velocity and buoyancy (cf. figure
\ref{diag:08}). In a turbulent flow, however, both APE and its
transport contain components that result from temporal
fluctuations, which we study in this section to explain the poor
agreement between the prediction and observation of the mean
buoyancy in figure \ref{fig:26}.

Due to the nonlinearity and convexity of $E_{a}(b)$, fluctuations
$b'=b-\av{b}$ in the buoyancy field increase the mean APE
$\av{E}_{a}$. Following \citet{ScoAjfm2014a}, it is therefore
convenient to decompose the time-averaged APE into a function of
the mean buoyancy field $\av{E}_{a}^{m}=E_{a}(\av{b})$ in addition
with the APE arising from the fluctuations of the buoyancy
$\av{E}_{a}^{t}=\av{E}_{a}-\av{E}_{a}^{m}$, such that

\begin{equation}
  E_{a}'\equiv \underbrace{\int_{b_{*}}^{b}(z_{*}(\hat{b})-z)\rd\hat{b}}_{E_{a}}-\underbrace{\overline{\int_{b_{*}}^{b}(z_{*}(\hat{b})-z)\rd\hat{b}}}_{\av{E}_{a}^{m}+\av{E}_{a}^{t}},
\label{eq:Ea_decomp}
\end{equation}

\noindent defines temporal fluctuations in the APE whose mean
$\av{E_{a}'}$ vanishes. Consequently, the average vertical
transport of APE comprises three physically distinct processes, because

\begin{equation}
  \av{wE_{a}}=\av{w}\av{E}_{a}^{m}+\av{w}\av{E}_{a}^{t}+\av{w'E_{a}'}.
  \label{eq:wEa2}
\end{equation}

\noindent The first term on the right-hand side of equation
\eqref{eq:wEa2} corresponds to transport by the mean vertical
velocity $\av{w}$ of the APE associated
with the \emph{mean} buoyancy $\av{b}$. The second term
corresponds to transport by the mean vertical velocity of the
APE associated with \emph{fluctuations} in
the buoyancy, and the third term corresponds to transport by the
turbulent velocity field. In general, fluctuations in the
background state defined by $z_{*}$ would lead to additional
terms. For further details the reader is referred to
\citet{ScoAjfm2014a}, who do not assume a steady background
buoyancy field \emph{a priori}.

Figure \ref{fig:25} displays the observed APE transport with
respect to the similarity coordinate $r/(r_{m}Z)$ at heights
$z=0.3\ (a,b)$, $z=0.1\ (c,d)$ and $z=0.0\ (e,f)$ for aspect
ratios $S=1$ $(a,c,e)$ and $S=4/3$ $(b,d,f)$. The most significant
feature of APE transport that can be seen in figure \ref{fig:25}
is the dominant contribution from the advection of buoyancy
fluctuations by the mean flow $\av{w}\av{E}_{a}^{t}$. In contrast,
the APE flux resulting from mean profiles of velocity and buoyancy
$\av{w}\av{E}_{a}^{m}$ is relatively insignificant, particularly
for large values of $z$.

A second notable feature of figure \ref{fig:25} is that the
profiles of the total APE flux $\av{wE_{a}}$ are approximately
self-similar, when scaled according to the theoretical model derived in
\S\ref{sec:similarity}. However, the components \eqref{eq:wEa2}
comprising $\av{wE_{a}}$ are not individually self-similar,
because their relative amplitude and shape changes with respect to
$z$. Indeed, self-similar profiles of mean velocity and mean
buoyancy will not produce self-similarity in
$\av{w}\av{E}_{a}^{m}$; according to \eqref{eq:Ea_layered} the
relative contribution of positively and negatively buoyant parts
of the jet to $\av{w}\av{E}_{a}^{m}$ depend on $z$ in different
ways, which violates self-similarity. Some evidence of this can be
found in figure \ref{fig:25}, which shows that the profile of
$\av{w}\av{E}_{a}^{m}$ becomes flatter as $z$ increases and one
moves further from the interface.

We conclude that the heuristic arguments leading the scaling
derived in \S\ref{sec:ape_model} should not be interpreted
literally in terms of mean-flow profiles. In contrast with
classical plume theory, from which fluctuating quantities are
typically neglected, the fluctuating buoyancy field plays a
dominant role in this flow's energetics via the flux
$\av{w}\av{E}_{a}^{t}$. Whilst the notion that parcels of
relatively dense fluid are dragged into the upper layer
(cf. figure \ref{diag:08}) is qualitatively correct, the process
should be understood as leading to temporal fluctuations in the
buoyancy field that are advected by the mean flow. In this regard
it is particularly interesting that the flow organises itself to
produce self-similarity in $\av{wE_{a}}$, in spite of the evolving
differences between the relative contributions from its
constituent parts \eqref{eq:wEa2}.

\section{Conclusions}
\label{sec:conclusions}

We have analysed the flow driven by equal and opposite point
sources of buoyancy in closed domains. In \S\ref{sec:mean} we
focused on the structure of the mean flow and buoyancy, whereas in
\S\ref{sec:flow_energetics} we focused on quantities pertaining to
the system's energetics. These contrasting perspectives
highlighted aspects of the problem that do and do not entail an
explicit mathematical dependence on an entrainment coefficient,
respectively. Unlike the strength of the resulting circulation and
buoyancy differences studied in \S\ref{sec:mean}, the vertical
distribution of BPE production in \S\ref{sec:flow_energetics}
(which was approximately equal to APE production in the cases we
considered) was predicted without recourse to an entrainment
coefficient. In this regard, entrainment determines a
`constitutive' relation between buoyancy (force) and volume flux
(flow) that is not visible from the perspective of energetics when
the power input to the system is fixed. An analogy in the context
of turbulent Rayleigh-B\'{e}nard convection would be the
relationship between the Nusselt number, buoyancy variance
dissipation and the flow's energetics resulting from fixed
buoyancy flux boundary conditions \citep[cf.][]{HugGjfm2013a}.

In the first part of this study we observed a stable two-layer
stratification and three distinct types of steady-state
circulation between the plumes on domains of sufficiently large
aspect ratio. The primary and largest circulation cell extended
over the full depth of the domain and corresponded to the
transport of fluid between the layers via turbulent entrainment
into the plumes. In each layer a secondary circulation cell,
driven by entrainment into the jet-like flows adjacent to each
plume, was observed. Between each secondary circulation, we
observed a third circulation cell of relatively small vertical
extent around the domain's mid-plane. We expect an understanding
of such flow structures to be relevant in predicting the transport
of airborne pollutants in confined environments. In this regard,
we note that the single circulation cell found in domains of
relatively small aspect ratio ($S\leq 15/24$) can bifurcate to
produce secondary circulation cells if the aspect ratio is
increased (see, for example, figure \ref{fig:01}). This
observation suggests that assumptions about the mean or
steady-state flow structure for a particular problem should be
made with caution.

We compared two estimates of an effective entrainment coefficient
for the system: one calculated from the volume flux at the
domains' mid-plane and a second corresponding to the observed
buoyancy difference between layers of constant buoyancy. For
domains of relatively small aspect ratio, the two estimations were
significantly different, due to the stratification of each layer and
the effects of background turbulence. For domains of relatively
large aspect ratio the two estimations coincided, but yielded an
effective entrainment coefficient of $0.2$, which is notably
higher than the typical value of $0.12$ corresponding to an
unconfined, non-interacting plume from a point source
\citep{ReeMjfm2015a}. A higher entrainment coefficient has
implications for the design of ventilation systems that utilise
stack-driven displacement ventilation, because the prediction of
the height of the occupied layer depends on the assumed value for
the entrainment coefficient.  Whilst the use of an `effective'
virtual source \citep[see e.g.][]{LinPjfm1990a, HunGjfm2001b} can
account for the relatively large entrainment coefficient, we would
expect its position to depend on the intensity of background
turbulence based on the trend exhibited in figure
\ref{fig:entrainment_rate} of this study.

In the second part of this study we examined and modelled the
system's energetics without invoking a closure to account for
entrainment. Consistent with \citet{HugGjfm2013a} and
\citet{GayBprl2013a}, we find a global mixing efficiency for this
convective flow of $1/2$. Our results are consistent with
\citet{WykMarx2018a}, who demonstrate that the mixing efficiency
of an emptying filling box is proportional to the relative depth
of the upper layer. We extended their analysis by developing an
analytical solution for Gaussian plume profiles. The production of
BPE and dissipation of APE was consequently found to be three
times larger in the layer containing the plume than the layer
containing the jet, whereas the viscous dissipation of kinetic
energy was divided equally between the two layers. In relation to
our observations concerning entrainment, an understanding the
system's energetics provides a logical step towards modelling the
effect that background turbulence might have on plumes in confined
environments.

We have examined confined plumes using both classical plume theory
and an energetics framework. Consistent with the energetics that
underpin Rayleigh-B\'{e}nard convection \citep{HugGjfm2013a}, the
dissipation of buoyancy variance which can be mathematically
related to an entrainment coefficient \citep{CraJjfm2017a}, does
not play a direct role in this system's energy budgets. We
contrast this situation, for \emph{closed} domains, with the
result for \emph{open} domains obtained by
\citet{WykMarx2018a}. There, an expression for the mixing
efficiency of an emptying filling box was found to depend on the
interface height, which, due to the box's connection with an
exterior, necessarily depends on the entrainment of volume into
the plume.

As a first step towards the use of local APE frameworks to
understand the physics behind entrainment, our hope is that the
volume- and plane-averaged perspective developed in
\S\ref{sec:flow_energetics} provides a useful point of reference
and link with existing work in the context of plume
modelling. Progress in this respect might come from closer
scrutiny of the radial dependence of APE dissipation in a plume,
particularly in the vicinity of the turbulent/non-turbulent
interface.

\section{Acknowledgements}

J.C. gratefully acknowledges an Imperial College Junior Research
Fellowship. Computational resources for this work came from an
EPSRC ARCHER Leadership grant, the `Cambridge Service for Data
Driven Discovery' (CSD3, \url{http://csd3.cam.ac.uk}) system
operated by the University of Cambridge Research Computing Service
(\url{http://www.hpc.cam.ac.uk}) funded by EPSRC Tier-2 capital
grant EP/P020259/1 and the High Performance Computing facility at
Imperial College London. We would also like to acknowledge helpful
suggestions from four anonymous referees, particularly those
concerning local APE frameworks.

\appendix

\section{Evaluation of the APE flux}
\label{sec:beta}

Following classical plume theory under the assumption of Gaussian
profiles for buoyancy and velocity, we assume that below the
interface the dimensionless velocity and buoyancy profiles can be
described by the functions

\refstepcounter{equation}
$$
  b = 4\,b_{m}\exp\left(-2\eta^{2}\right)Z^{-5/3}-b_{m},\quad
  w = 2\,w_{m}\exp\left(-2\eta^{2}\right)Z^{-1/3},
\eqno{(\theequation{\mathit{a},\mathit{b}})}\label{eq:wb}
$$

\noindent where $Z\equiv (z + \zeta)/\zeta$ is the distance from the
source rescaled by the interface height, 
$\eta=r/(r_{m}Z)$ is a similarity coordinate based on the
radial distance from the axis of the plume (see figure
\ref{diag:01}), $r_{m}=6\alpha\zeta/5$ is the radius of the plume at
the interface and

\begin{equation}
  w_{m}=\frac{5}{6\alpha}\left(\frac{9\alpha}{10\pi}\right)^{1/3}\zeta^{-1/3},
\end{equation}

\noindent is a characteristic velocity at the interface, such that $Q_{m}=\pi w_{m}r_{m}^{2}$.

Assuming a step change from $-b_{m}$ to $b_{m}$ in the ambient
buoyancy across the interface, the buoyancy just above the
interface is:

\begin{equation}
  b = \left( 4b_{m}\exp\left( -2\eta^{2}\right)-2b_{m}\right)Z^{-c-1}+b_{m},
\label{eq:bA}
\end{equation}

\noindent where $c=10/3$, as derived in \S\ref{sec:similarity}.
The velocity in the upper layer is assumed to scale according to a
turbulent jet:

\begin{equation}
  w = 2\,w_{m}\exp\left( -2\eta^{2}\right)Z^{-1}.
\label{eq:wA}
\end{equation}

\noindent As required, the profiles \eqref{eq:wb}, \eqref{eq:bA}
and \eqref{eq:wA} are such that the relative buoyancy flux

\begin{equation}
  2\pi\int_{0}^{\infty}w(b-b_{*})r\mathrm{d}r=
  \begin{cases}
    1,&\quad\quad z<0,\ Z<1,\ b_{*}=-b_{m},\\
    0,&\quad\quad z>0,\ Z>1,\ b_{*}=b_{m}.
    \end{cases}
\end{equation}

\subsection{Evaluation of $\hav{wE_{a}}$ below the interface $Z<1$, $z<0$}

To evaluate the integral of the APE flux $wE_{a}$ below the
interface we substitute \eqref{eq:wb} into \eqref{eq:Ea_layered}
and integrate over the area of a single plume. To account for
$z_{*}(b)$, which \eqref{eq:zs} indicates is discontinuous with
respect to $b$, it is useful to recognise that $b=b_{m}$ in
\specialeqref{eq:wb}{a} when
$\eta=\eta_{0}=\sqrt{\log(2Z^{-5/3})/2}$ and split the resulting
integral into two parts. Noting
that $2\pi\,w_{m}b_{m}r_{m}^{2}=1$ is the buoyancy flux,

\begin{equation}
  \begin{aligned}
2\pi r_{m}^{2}Z^{2}\int_{0}^{\eta_{0}}wE_{a}\eta\mathrm{d}\eta &=1-\zeta\,Z-(1-\zeta)Z^{5/3}+\frac{1-2\zeta}{4}Z^{10/3}+\frac{\zeta}{4}Z^{13/3},\\
2\pi r_{m}^{2}Z^{2}\int_{\eta_{0}}^{\infty}wE_{a}\eta\mathrm{d}\eta &=\frac{\zeta}{4}Z^{10/3}-\frac{\zeta}{4}Z^{13/3},
  \label{eq:wEa_integral}
  \end{aligned}
\end{equation}
\noindent such that 
\begin{equation}
  \hav{wE_{a}}=1-\zeta\,Z-(1-\zeta)Z^{5/3}+\frac{1-\zeta}{4}Z^{10/3}.
\label{eq:wEa_below}
\end{equation}

\subsection{Evaluation of $\hav{wE_{a}}$ above the interface $Z>1$, $z>0$}

\noindent As explained in \S\ref{sec:similarity}, above the
interface

\begin{equation}
  \hav{wE_{a}}=(1-\zeta)\frac{\hav{|\Phi_{z}|}}{2}=(1-\zeta)\frac{\hav{|w(b-b_{m})|}}{2}.
\end{equation}

\noindent The integral of $|\Phi_{z}|$ can be evaluated by
integrating $w(b_{m}-b)>0$ over the shell in which
$b<b_{m}$. According to \eqref{eq:bA}, $\eta_{0}=\sqrt{\log(2)/2}$
defines the point at which $b=b_{m}$ when $Z\geq 1$; hence, noting
again that $2\pi\,w_{m}b_{m}r_{m}^{2}=1$,

\begin{equation}
  \frac{\hav{|\Phi_{z}|}}{2}=\frac{1}{Z^{c}}\int^{\infty}_{\eta_{0}}\left(\exp\left(-2\eta^{2}\right)-2\exp\left(-4\eta^{2}\right)\right)\eta\mathrm{d}\eta = \dfrac{1}{4Z^{c}}.
  \label{eq:Phiz_above}
\end{equation}

\noindent At the interface $Z=1$ and
$(1-\zeta)\hav{|\Phi_{z}|}/2=1/8$, which is consistent with
\eqref{eq:wEa_below}. As $\beta$ is defined as
$\hav{|\Phi_{z}|}/2$ evaluated at $Z = 1$, \eqref{eq:Phiz_above}
shows that $\beta = 1/4$ for a plume with Gaussian velocity and
buoyancy profiles.

\section{Validation}
\label{sec:valid}

To validate our findings, we ran each of the simulations reported
in the main text with a reduced grid resolution, as described in
\S\ref{sec:grid}. In \S\ref{sec:rey}, we describe simulations that
were run to check the sensitivity of the flow structures discussed
in \S\ref{sec:mean} with respect to a reduction in the Reynolds number.

\subsection{Grid resolution}
\label{sec:grid}

To verify the independence of our results to grid resolution we
ran each of the simulations listed in table \ref{tab:02} using
half as many computational cells in each spatial
direction. Specifically, we employed $384$ cells in the vertical
($z$) direction and $384, 438, 480, 512, 768, 1024$ cells in the
horizontal directions for quadrant aspect ratios
$1/2, \, 7/12, \, 15/24, \, 2/3, \, 1$, and $4/3$, respectively.

\begin{figure}
  \begin{center}
    \includegraphics[scale=1]{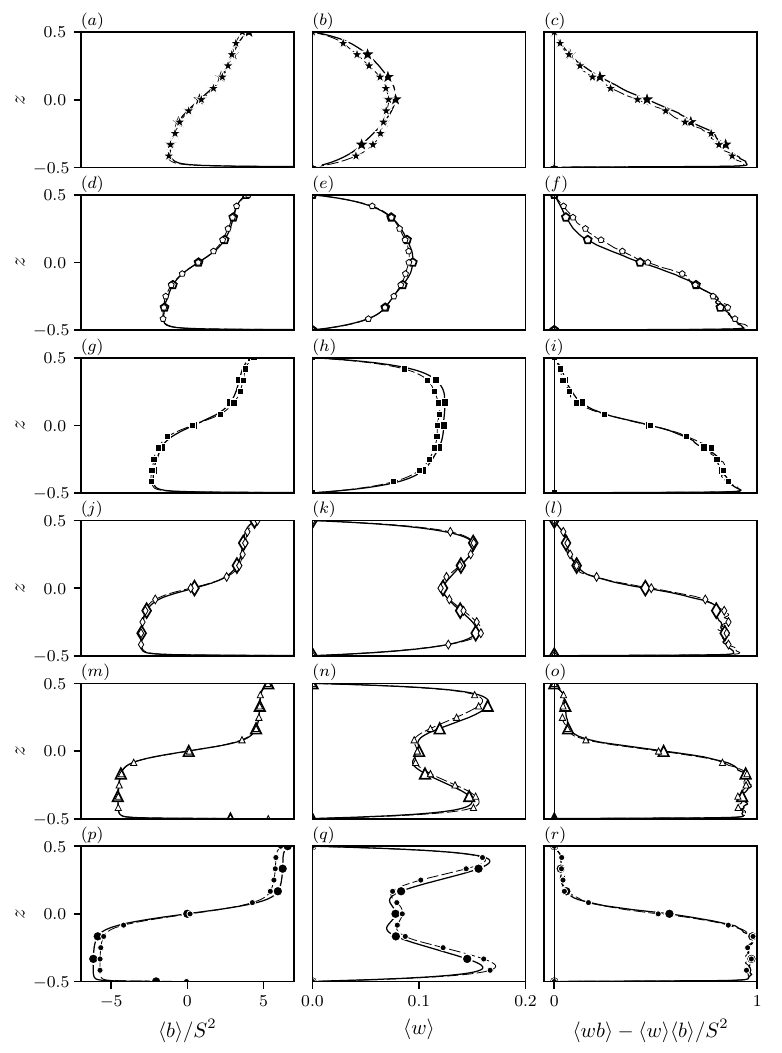}
  \end{center}
  \caption{Grid resolution study. Horizontally- and time-averaged
    buoyancy (first column), with quadrant integrals of volume flux
    (second column) and relative buoyancy flux (third column) for
    aspect ratios $S=1/2, \, 7/12, \, 15/24, \, 2/3, \, 1$, and
    $4/3$, from top row to bottom row, respectively.  The symbols
    correspond to those defined in table \ref{tab:02} and used in
    figure \ref{fig:01}. The dashed lines and relatively small
    symbols correspond to results obtained at the full grid
    resolutions reported in table \ref{tab:02}. The solid lines
    and relatively large symbols correspond to results obtained
    from grids employing half as many grid cells as those reported
    in table \ref{tab:02} in each spatial direction.}
  \label{fig:31}
\end{figure}

The quantities displayed in figure \ref{fig:31}, corresponding to
those of figure \ref{fig:01}, indicate a good agreement between
the results obtained from each of the two computational grids. The
coarser grids reproduce the difference in buoyancy between the
upper and lower layers (figure \ref{fig:31}$(a,d,g,j,m,p)$), in
addition to the primary, secondary and tertiary circulation cells
discussed in \S\ref{sec:mean}. With regards to the latter, we note
that the results of the original simulations were not used as
initial conditions for the coarser grids, and therefore regard the
existence of the circulation patterns reported in \S\ref{sec:mean}
as robust for the parameters used in this study.

\begin{figure}
  \begin{center}
    \includegraphics[scale=1]{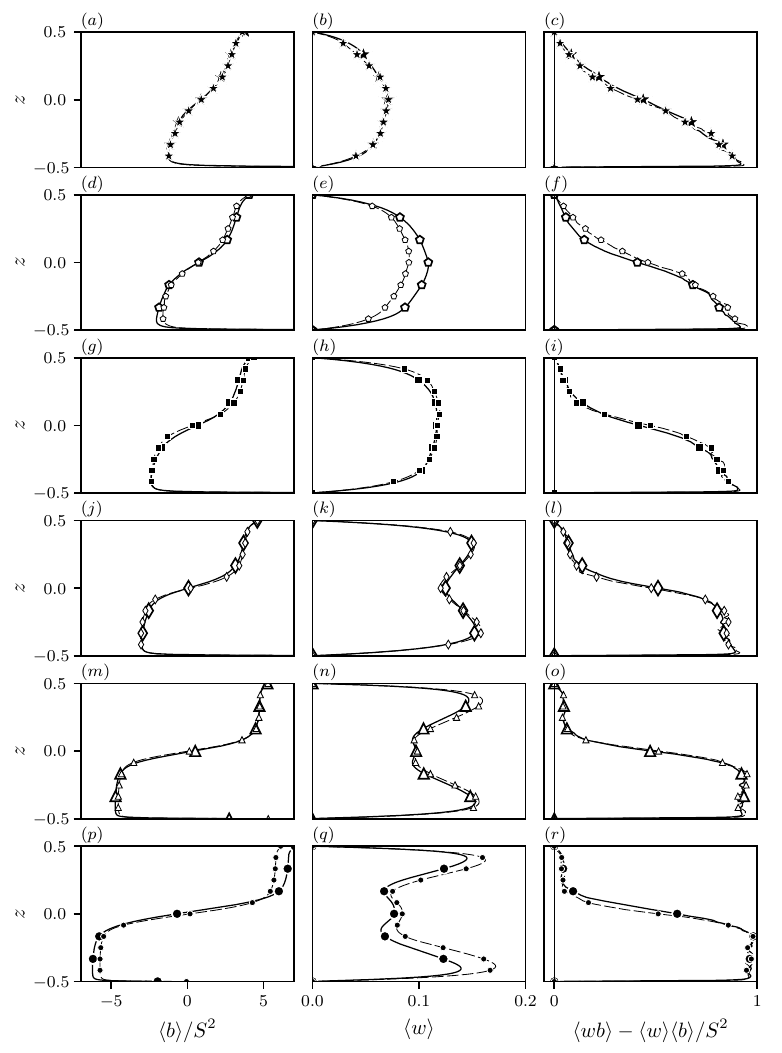}
  \end{center}
  \caption{Reynolds number study. Horizontally- and time-averaged
    buoyancy (first column), with quadrant integrals of volume
    flux (second column), and relative buoyancy flux (third
    column) for aspect ratios
    $S=1/2, \, 7/12, \, 15/24, \, 2/3, \, 1$, and $4/3$, from top
    row to bottom row, respectively. The comparison is between
    results for $Re=4185$ (dashed lines and small symbols) using
    the grids reported in table \ref{tab:02} and those for
    $Re=2929$ (solid lines and large symbols) using coarse grids,
    comprising half as many points as those in table \ref{tab:02}
    in each spatial direction.}
  \label{fig:32}
\end{figure}

\subsection{Reynolds number}
\label{sec:rey}

To observe the local effect of changes in Reynolds number to the
simulation results, we reduced the Reynolds number from $4185$ to
$2929$, whilst employing the coarser computational grid described
in \S\ref{sec:grid}. The results shown in figure \ref{fig:32}
suggest that the reduction in Reynolds does not cause a
qualitative change in our findings regarding the internal
temperature structure and circulation patterns. In particular, at
$Re=2929$ the plumes are turbulent and, based on there being no
observed change in the temperature difference between upper and
lower layers in figure \ref{fig:32}$(p)$, entrain at approximately
the same rate as when $Re=4185$. The results suggest that the
existence of primary, secondary and tertiary circulation cells are
robust to relatively small changes in Reynolds number,
notwithstanding differences in the maximum volume flux seen in
panels $(e, S=7/12)$ and $(q, S=4/3)$. In the case of
$(e, S=7/12)$, the quadrant volume flux depends sensitively on
aspect ratio and on the slowly-varying organisation of the flow,
which makes the precise determination of Reynolds number
sensitivity difficult. Therefore, without data corresponding to
additional aspect ratios for a larger range of Reynolds numbers,
we note that the precise aspect ratios at which bifurcations in
the mean flow between quadrants occur could depend on Reynolds
number.

\bibliographystyle{jfm}
\bibliography{main}

\end{document}